\documentclass[10pt]{book}
\usepackage{array}
\usepackage{amsmath,amssymb,amsthm}

\usepackage{graphicx}
\usepackage{url}

\newtheorem{thm}{Theorem}[chapter]

\newtheorem{prop}[thm]{Proposition}

\newtheorem{rmk}[thm]{Remark}
\newtheorem{ex}[thm]{Example}

\begin{document}
\renewcommand{\baselinestretch}{1.2}
%\lhead[\fancyplain{} \leftmark]{}
%\chead[]{}
%\rhead[]{\fancyplain{}\rightmark}
%\cfoot{}
%\headrulewidth=0pt
\markright{
%\hbox{\footnotesize\rm Statistica Sinica
%{\footnotesize\bf ??}(200?), 000-000}\hfill
}
\markboth{\hfill{\footnotesize\rm N.~Beerenwinkel, L.~Pachter, and B.~Sturmfels
}\hfill}
{\hfill {\footnotesize\rm EPISTASIS AND SHAPES OF FITNESS LANDSCAPES} \hfill}
\renewcommand{\thefootnote}{}
$\ $\par
\fontsize{10.95}{14pt plus.8pt minus .6pt}\selectfont
\vspace{0.8pc}
\centerline{\large\bf EPISTASIS AND SHAPES OF FITNESS LANDSCAPES}
%\vspace{2pt}
%\centerline{\large\bf IF A SECOND LINE IS NEEDED}
\vspace{.4cm}
\centerline{Niko Beerenwinkel, Lior Pachter, and Bernd Sturmfels}
\vspace{.4cm}
\centerline{\it Department of Mathematics, University of California at Berkeley}
\vspace{.55cm}
\fontsize{9}{11.5pt plus.8pt minus .6pt}\selectfont

\begin{quotation}
\noindent {\it Abstract:}
The relationship between the shape of a fitness landscape and the
underlying gene interactions, or epistasis, has been extensively studied
in the two-locus case. Gene interactions among multiple loci 
are usually reduced to two-way interactions. 
We present a geometric theory of shapes
of fitness landscapes for multiple loci. 
A central concept is the genotope, which is the convex hull of
all possible allele frequencies in
populations. Triangulations of the genotope correspond to different 
shapes of fitness landscapes and reveal all the gene interactions. 
The theory is applied to fitness 
data from HIV and Drosophila melanogaster.
In both cases,
our findings refine earlier analyses and reveal 
previously undetected gene interactions. 
\par

\vspace{9pt}
\par
\end{quotation}\par

\fontsize{10.95}{14pt plus.8pt minus .6pt}\selectfont

%%%%%%%%%%%%%%%%%%%%%%%%%%%%%%%%%%%%%%%%%%%%%%%%%%%%%%%%%%%%%%%%%%%%%%%%%
\setcounter{chapter}{1}
\setcounter{equation}{0} %-1
\noindent {\bf 1. Introduction}
\smallskip

The term ``epistasis'' was coined by
Bateson (1909) to describe the interactions among
individual genes. The concept was introduced in the setting
of Mendelian genetics, where epistasis gives rise to 
distorted Mendelian ratios of genotypes. In the context of
statistical genetics, epistasis was originally called ``epistacy'' 
by Fisher (1918). Here, it arises when mapping discrete 
genotypes to continuous traits and
refers to contributions to the phenotype that are not linear 
in the average effects of the single genes. 
The phenotypic trait of an organism that drives 
the evolution of the population is the reproductive
fitness of individuals, i.e., the expected number of
offspring. For this trait, the genotype--phenotype
mapping is called a fitness landscape 
(Gavrilets, 2004; Wright, 1931), 
and epistasis is a property of that landscape.   
 
\begin{figure}
\centering
\includegraphics[width=\textwidth]{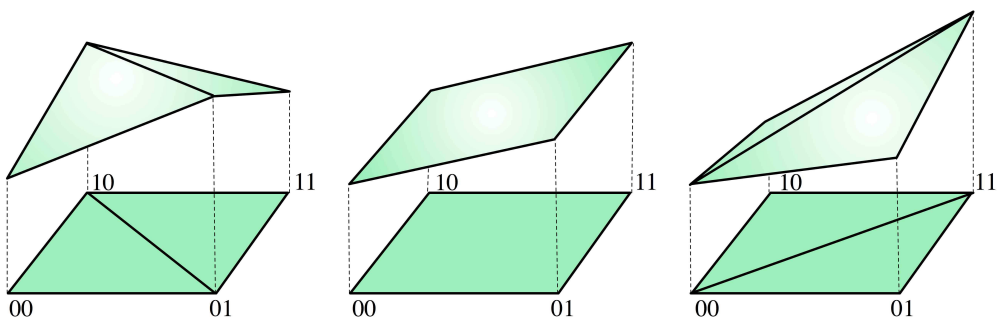}
\caption{
The possible shapes of fitness landscapes
on two biallelic loci. The genotope is the square.
Its two triangulations correspond to
negative and positive epistasis.}
\label{fig:updown}
\end{figure}

For a genetic system of two biallelic loci, there is only
one type of interaction, illustrated
by the landscapes in Figure~\ref{fig:updown},
and epistasis refers unambiguously to this interaction. 
However, the situation is more complex for more than two
loci, because new interaction patterns arise.
The current language for describing gene interactions 
does not reflect this diversity (Phillips, 1998).
Indeed, the common approach of investigating epistasis by 
analysis of variance (ANOVA) and expressing
epistatic effects as the residuals of a  
linear regression of fitness on the
genotypes (Cordell, 2002) does not account for the
intrinsic combinatorial structure of the landscapes that
generalize two-locus epistasis.

For a specified set of genotypes that may involve multiple 
loci, we characterize all possible interactions
among them. Our characterization is based on a 
geometric object, the {\em genotope}, which is the convex
hull of the possible allele frequency vectors. The
regular triangulations (De~Loera et al., 2006)
of the genotope encode the genotype interactions 
in the fitness landscape. The biological 
problem of studying genotype interactions for a fitness landscape is
thus equivalent to the combinatorial problem of finding the
shape of the fitness landscape, i.e., the triangulation of the
genotope that is induced by the fitness landscape. 

In Sections~2 to 4, the mathematical concepts are introduced
and illustrated in several examples. 
For simplicity, we focus primarily on the case of 
a population of haploid individuals 
(or, equivalently, of homozygous diploids). Nevertheless, our
concepts and algorithms are not limited to haploids,
and in Example \ref{diploids}
we briefly indicate the modifications necessary for diploids.
 
In Section~5, we discuss the 
three-locus two-allele system, which amounts to classifying the
$74$ triangulations of the 3-cube. Table~\ref{tab:74}
is the direct generalization of the list of 
possible shapes for
two biallelic loci, shown in Figure~\ref{fig:updown}.

In Sections~6 and 7, we apply
our method to two published fitness landscapes. Section~6 deals
with a biallelic three-locus system in HIV
(Segal et al., 2004) and emphasizes the
modeling of measurement error. Section~7 revisits the
five-locus system in Drosophila melanogaster studied by
Whitlock and Bourguet (2000).  
The shape of their  Drosophila
fitness landscape is a  triangulation 
 of the $5$-dimensional cube (the genotope) into $110$
simplices, listed in Figure~\ref{fig:simplices}.

Our discussion in Section~8 compares our approach with
other studies of fitness landscapes. It also raises the 
challenge of determining the {\em human genotope} 
and its biologically relevant triangulations
from suitable haplotype  data. \par

%%%%%%%%%%%%%%%%%%%%%%%%%%%%%%%%%%%%%%%%%%%%%%%%%%%%%%%%%%%%%%%%%%%%%%%%%
\bigskip
\setcounter{chapter}{2}
\setcounter{equation}{0} %-1
\noindent {\bf 2. Populations and the Genotope}
\smallskip

We fix a finite alphabet $\Sigma$ of size $l$, which labels the
$l$ different alleles at a genetic locus of interest.
The elements of $\Sigma$ may correspond to variants of a gene, 
to the nucleotides at a genome position 
($\Sigma= \{{\tt A}, {\tt C}, {\tt G}, {\tt T}\}$, $l=4$),
or to the amino acids at a codon position in a gene ($l=20$).
The biallelic case $(\Sigma = \{0,1\}, l=2)$ arises frequently in genomics,
where it is  known that
most SNPs (single nucleotide polymorphisms) 
have two types. 

The allele frequencies at a certain locus
define a probability distribution on the alphabet $\Sigma$. The set of
all probability distributions on $\Sigma$ 
 is identified with the $(l-1)$-dimensional standard simplex
\[
   \Delta_\Sigma \quad = \quad \bigl\{ (p_1,p_2, \ldots, p_l) \in [0,1]^l
\,\,     \colon \,\,p_1+ p_2+ \cdots + p_l = 1 \bigr\}.  
\]
We consider $n$ loci, all with the same alphabet $\Sigma$
of alleles, and we denote by $\Sigma^n$ the set of sequences of length 
$n$ over $\Sigma$. The elements of $\Sigma^n$ are identified with 
the vertices of the $n$-fold direct product of simplices
$\Delta_\Sigma^n = (\Delta_\Sigma)^n$. This
product is a convex polytope  (Ziegler, 1995)
of dimension $ln-n$ having $l^n$ vertices. In particular, in the binary case,
$\Delta_{\{0,1\}}^n$ is the standard $n$-dimensional cube.

A {\em genotype space} is any subset $\mathcal{G}$ of $\Sigma^n$.
For a given genotype space $\mathcal{G}$,  
let $\Pi_{\mathcal{G}}$ be the convex hull of all vertices in
$\Delta_{\Sigma}^n$ that are indexed by sequences in
$\mathcal{G}$. The set $\Pi_{\mathcal{G}}$ 
is a subpolytope of  $\Delta_{\Sigma}^n$.
We call $\Pi_{\mathcal{G}}$  the {\em genotope}.
A point $v$ in the genotope is an $n$-tuple of allele 
frequencies. To be precise, if $\,v = (v_1, \dots, v_n)\,
 \in \Pi_{\mathcal{G}} \subseteq \Delta^n_\Sigma\,$ then
$v_i \in \Delta_\Sigma$ represents the allele frequencies 
at locus $i$.

\begin{ex} \label{revisited}
{\bf (Genotype lattices)} \hfill \break \rm
In the study of directed evolution in (Beerenwinkel et al., 2006)
the alphabet is $\Sigma = \{0,1\}$ 
and the genotype space $\mathcal{G}$ is the distributive lattice
induced by an event poset $\mathcal{E}$ with $n$ elements.
The genotope $\Pi_\mathcal{G}$ is the {\em order polytope} of the 
poset $\mathcal{E}$. If the event poset $\mathcal{E}$ is empty then
$\mathcal{G} = \{0,1\}^n$, the genotope $\Pi_\mathcal{G}$
is the $n$-dimensional cube.
The case $n=2$ is Example \ref{tetrasquare}.
The case $n=3$ is discussed in detail in Section~4. 
For an example of a non-empty event poset consider
 $n =3$ and $\mathcal{E} = \{ 2 < 3\}$, meaning that the second event has to 
occur before the third event can happen. 
The induced genotype lattice has six genotypes,
\begin{equation}
\label{toblerone}
 \mathcal{G} \quad = \quad \{ \,
 000, \, 001, \, 011,\,  100, \, 101, \, 111\, \}, 
\end{equation}
and the corresponding genotope $\Pi_{\mathcal{G}}$ is a triangular prism.
\qed
\end{ex}

\begin{figure}
\centering
\includegraphics[scale=0.5]{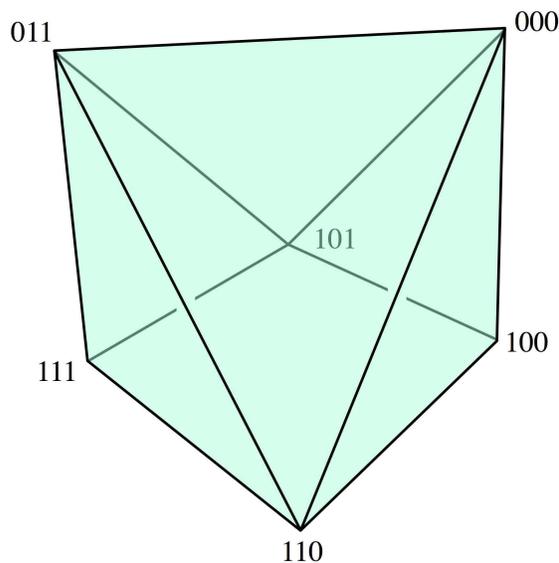}
\caption{
A three-dimensional genotope from the HIV data of Section 6.
}
\label{fig:gold}
\end{figure}

\begin{ex}   \label{threesix}
{\bf (A genotope from HIV fitness data)}  \hfill \break \rm
Consider the genotype space
\begin{equation}
\label{nottoblerone}
 \mathcal{G} \quad = \quad \{ \,
 000,\, 110, \, 011,\,  100, \, 101, \, 111\, \},
\end{equation}
which differs from the one in (\ref{toblerone})
only by the second genotype. This $\mathcal{G}$ does not form  
a distributive lattice, so it does not
arise in the setting of (Beerenwinkel et al., 2006).
The genotope $\Pi_{\mathcal{G}}$ has
six triangular faces and one square face (Figure \ref{fig:gold}).
This genotope appears 
in our analysis of the HIV data in Section~6. \qed
\end{ex}

By a {\em population} on $\mathcal{G}$ we shall mean 
any probability distribution $p$ on the set $\mathcal{G}$.
For any genotype $g \in \mathcal{G}$, the 
coordinate $p_g$ of $p$ represents
the fraction of the population that is of genotype $g$.
A population is a point in the {\em population simplex} 
$\Delta_{\mathcal{G}}$.
The population simplex and the genotope are related via the 
marginalization map $\rho$, which maps a population $p$ to its 
$n$-tuple of allele frequencies, 
\[
\rho \,\colon\, \Delta_{\mathcal{G}} \,\rightarrow \,
\Pi_\mathcal{G} \subset \Delta_{\Sigma}^n \, , \quad
\left( p_{\sigma_1 \dots \sigma_n} \right)_{\sigma \in \Sigma^n}
\, \mapsto\,  \left( \left( \sum_{\sigma:\sigma_i = \tau} 
 p_{\sigma_1 \dots \sigma_n}
 \right)_{\tau \in \Sigma} \, \right)_{\!\! i=1,\dots,n}. 
\]
If the population $p$ consists of a single genotype $g$,
then $p$ is the unit vector whose coordinates are $p_g= 1$ and
$p_h  = 0$ for all $h \in \mathcal{G}\backslash \{g\}$.
Its allele frequency vector $\rho(g)$ is a list of $n$ unit vectors,
each of length $l$.
The unit vector $\rho(g)_i$ is the vertex of the simplex
$\Delta_{\Sigma}$  indexed by the $i$-th allele
of the genotype $g$. Thus $\rho(g)$ is a vertex of the
genotope $\Pi_\mathcal{G}$, and all vertices arise in this manner.
Since the marginalization map $\rho$ is linear, we conclude:

\begin{prop}
The genotope $\Pi_\mathcal{G}$ equals the set of all possible
$n$-tuples of allele frequencies that may arise from a population
on the genotype space $\mathcal{G}$.
\end{prop}

\begin{figure}
\centering
\includegraphics[scale=0.6]{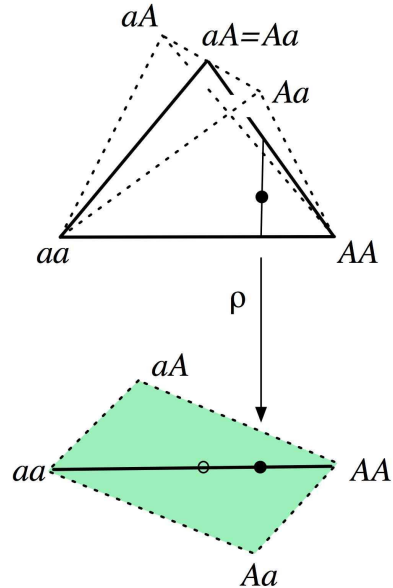}
\caption{Population simplex and genotope for one diploid biallelic
  locus.}
\label{fig:hardy}
\end{figure}

\begin{ex} \label{tetrasquare}
{\bf (Tetrahedron maps onto square)} \hfill \break \rm
Let $n=2$, $\Sigma = \{0, 1\}$, and consider the genotype space
 $\mathcal{G} =  \{00, 01, 10, 11\}$.
The set $\Delta_{\mathcal{G}}$ of probability distributions on 
$\mathcal{G}$ is a tetrahedron in 4-space,
\[
   \Delta_{\mathcal{G}}
\quad = \quad \left\{ (p_{00},\, p_{01},\, p_{10},\, p_{11}) \in [0,1]^4 
     \: \mid \: p_{00} + p_{01} + p_{10} + p_{11} = 1 \right\}.
\]
The genotope $\Pi_{\mathcal{G}}$ is the square $\Delta_\Sigma \times \Delta_\Sigma$.
A population $p$ is a point in the tetrahedron whose
allele frequencies are given by 
the marginalization map
\[
   \rho(p_{00}, p_{01}, p_{10}, p_{11}) \quad = \quad
     \left( (p_{00} + p_{01},\, p_{10} + p_{11}),\,
     (p_{00} + p_{10},\, p_{01} + p_{11}) \right).
\]
If we identify the square $\Pi_{\mathcal{G}} = \Delta^2_\Sigma$ with the convex hull 
of the four points
$(0,0)$, $(0,1)$, $(1,0)$, $(1,1)$ in the plane, then
the marginalization map
is the following projection
(see Figure~\ref{fig:hardy})
 from the tetrahedron onto the square:
\[
   \rho(p_{00}, p_{01}, p_{10}, p_{11}) \quad = \quad
     (p_{10} + p_{11},\, p_{01} + p_{11}).
\]
The two coordinates are the frequencies at the two loci
of the allele ``1''.
\qed
\end{ex}

\begin{ex} \label{diploids}{\bf (From haploids to diploids)} \hfill \break \rm
The mathematical set-up introduced so far is for haploids
only. It can be extended to diploids
using  either of the following two equivalent approaches.
 The allele frequencies of diploid
populations can be modeled by scaling each simplex in
$\Delta_{\Sigma}^n$ by a factor of two, so that the genotope sits inside
$(2 \Delta_\Sigma)^n$.
Alternatively, we could replace $n$ by $2n$ and  then restrict 
to symmetric populations in $\Delta_\Sigma^{2n}$.

To illustrate both approaches, we
consider a single biallelic 
locus ($l=2$, $n=1$) with  genotypes $aa, AA, aA, Aa$,
where we identify $Aa$ with $aA$. The population simplex is a triangle, 
and the genotope $2\Delta_\Sigma$ is a line segment  whose end points
are labeled $aa$ and $AA$ and whose midpoint is labeled
by  $Aa = aA$. This picture arises from 
$\Delta_{{\Sigma}^2} \rightarrow \Delta_{\Sigma}^2$ 
by intersecting with a plane. The genotope is 
the diagonal segment in the square of Example~\ref{tetrasquare},
and the population triangle is a cross-section of the tetrahedron,
depicted  in Figure~\ref{fig:hardy}. A point in the population triangle
has three coordinates $(p_{aa},\, p_{aA},\, p_{AA})$ where
 $p_{aa}$ is the frequency of genotype $aa$, 
$p_{AA}$ that of genotype $AA$, and $2p_{aA}$ that of genotype $Aa = aA$. 
%  
%  For two biallelic loci $(l=n=2)$, there are
%  nine possible genotypes,
%  $aabb$, $aAbb$, $\ldots$, $AABB$, and the genotope
%  $(2\Delta_\Sigma)^n$  is twice the unit square. 
\qed
\end{ex}

Returning to the general haploid model, we give
an interpretation of the fiber of a point in the genotope
$\Pi_{\mathcal{G}}$ under the the marginalization map $\rho$.
The fiber over $v \in \Pi_{\mathcal{G}}$ is
the following polytope inside the population simplex:
\begin{equation}
\label{fiber}
 \rho^{-1}(v) \quad = \quad
\bigl\{\, p \in \Delta_{\mathcal{G}} \,:\,
\rho(p) \, = \, v \,\bigr\}. 
\end{equation}

\begin{rmk}
If $v$ is any $n$-tuple of allele frequencies then
its fiber $\rho^{-1}(v)$ consists of
all populations $p$ which have the specified allele frequencies $v$.
\end{rmk}

If all coordinates of $v$ are non-zero, then   $\, \rho^{-1}(v) \,$
is  a polytope of  dimension 
\begin{equation}
\label{codimension}
 c(\mathcal{G}) \quad = \quad {\rm dim}(\Delta_{\mathcal{G}}) - 
{\rm dim}(\Pi_{\mathcal{G}}) \quad = \quad |\mathcal{G}|  
- {\rm dim}(\Pi_{\mathcal{G}}) - 1. 
\end{equation}
For the full genotype space $\mathcal{G} =
\Sigma^n$ we have $c(\mathcal{G}) = l^n-n-1$.
In particular, in  Example \ref{tetrasquare}
the dimension is one, and in Example \ref{threesix}
the dimension is two.

Since the fibers (\ref{fiber}) over the genotope $\Pi_\mathcal{G}$ characterize populations with
constant allele frequency vectors, we can restrict our attention
to these fibers whenever the allele frequencies are fixed.
For example, if evolution acts on the genotype space by recombination, 
but without mutation and selection, then the allele frequencies of the
population are constant for every generation. 
Hence, such evolutionary dynamics can be modeled by a
dynamical system on $\, \rho^{-1}(v) $.

Our geometric theory is adapted to the fact that, 
in biological systems,
the set $\mathcal{G}$ of observed genotypes
is usually significantly smaller than the number $l^n$
of possible genotypes. Specifically,
for binary data $(l=2)$ on many loci (say, $n \geq 20$),
 the genotope is never an $n$-cube, and its dimension
is smaller than $n$ due phenomena such as linkage.
The frequently heard assertion that dimensions
of  genotype spaces and fitness landscapes
increase exponentially in the sequence length
is thus misleading. Even for large data sets, the 
complexity of the genotope $\Pi_\mathcal{G}$ 
can be expected to be in the tractable range 
of polyhedral algorithms.
\par

%%%%%%%%%%%%%%%%%%%%%%%%%%%%%%%%%%%%%%%%%%%%%%%%%%%%%%%%%%%%%%%%%%%%%%%%%
\bigskip
\setcounter{chapter}{3}
\setcounter{equation}{0} %-1
\noindent {\bf 3. Fitness Landscapes and Interaction Coordinates}
\smallskip

A {\em fitness landscape} on a genotype space $\mathcal{G}$
is a function $w \colon \mathcal{G} \rightarrow \mathbb{R}$.
Each coordinate $w_g$ of $w$ denotes the logarithm of the 
reproductive fitness of genotype~$g$. 
The space of all fitness landscapes is the
$|\mathcal{G}|$-dimensional $\mathbb{R}$-vector space 
$\mathbb{R}^\mathcal{G}$. 
%For a fitness landscape 
%$w \in \mathbb{R}^\mathcal{G}$ and a genotype $g \in \mathcal{G}$, 

In the study of gene interactions one considers linear combinations of 
the measured coordinates $w_g$. 
These linear combinations express epistatic effects.
Certain collections of such linear forms
play the role of interaction coordinates
 on $\mathbb{R}^\mathcal{G}$.
 Both the sign and the magnitude
of these interaction coordinates are of 
interest when examining the 
fitness landscape $w$ of a biological system.

\begin{ex}  \label{twotriangles}
{\bf (Two triangles over a square)} \hfill \break \rm 
As in  Example~\ref{tetrasquare}, let
$\mathcal{G} = \{0,1\}^2 = \{00,01,10,11\}$, so the
 genotope $\Pi_\mathcal{G}$ is a square. A fitness
landscape $w$ is specified by the four numbers
$w_{00}$, $w_{01}$, $w_{10}$, and $w_{11}$, which 
we visualize as heights over the vertices of the square
$\Pi_\mathcal{G}$.
The interaction between the two loci  is
measured by the {\em epistasis}
\[
   u \quad = \quad  w_{00} + w_{11} - w_{01} - w_{10}.
\]
This defines three equivalence classes in the space of 
fitness landscapes according to whether the epistasis $u$ is positive,
negative, or zero.  This trichotomy is depicted in
 Figure~\ref{fig:updown}.
This geometric view of genotype interaction 
leads us (in Section~4) to a natural concept of shapes of
fitness landscapes for $n > 2$ loci. \qed
\end{ex}

Let $\mathcal{L}_{\mathcal{G}}$ be the subspace of
 $\mathbb{R}^\mathcal{G}$ 
consisting of all fitness landscapes $w$ which
have no interaction. Mathematically, $w$ is in $\mathcal{L}_{\mathcal{G}}$
if there is an affine-linear function on
the genotope $\Pi_\mathcal{G}$ whose values
at the vertices are the  $w_g$.
We define the {\em interaction space} $\mathcal{I}_\mathcal{G}$
as the vector space dual to the quotient of
$\mathbb{R}^\mathcal{G}$ modulo~$\mathcal{L}_{\mathcal{G}}$:
$$ \mathcal{I}_\mathcal{G} \quad := \quad
(\mathbb{R}^\mathcal{G}/\mathcal{L}_{\mathcal{G}})^* . $$
An element of the interaction space $\mathcal{I}_\mathcal{G}$ is
a linear form in the unknowns  $w_g$ which 
vanishes identically on the subspace  $\mathcal{L}_{\mathcal{G}}$.
In Example~\ref{tetrasquare}, the space
$\mathbb{R}^\mathcal{G}$ is four-dimensional,
its subspace
$\mathcal{L}_{\mathcal{G}}$ is three-dimensional,
and $\, \mathcal{I}_\mathcal{G} \,$
  is the one-dimensional space spanned by
  $\,  u \, = \,  w_{00} + w_{11} - w_{01} - w_{10}$.
In general, the dimension of the interaction space
$\, \mathcal{I}_\mathcal{G} \,$ is the quantity
$ c(\mathcal{G})$ defined in Equation~\ref{codimension}.

The interaction space $ \mathcal{I}_\mathcal{G}$
is spanned (redundantly) by a canonical set
of linear forms which are known as the {\em circuits}.
These are the linear forms whose
support (i.e., the $w_g$ which appear
with non-zero coefficient) is non-empty
but minimal with respect to inclusion.
The circuits in  $ \mathcal{I}_\mathcal{G}$
are unique up to scaling. Their number is
usually larger than $c(\mathcal{G})$ but it is
bounded above by 
$\,\binom{|\mathcal{G}|}{c(\mathcal{G})-1} $.

To see this upper bound, we note that
$\mathcal{L}_{\mathcal{G}}$ has dimension
$\,|\mathcal{G}|- c(\mathcal{G})$. The circuits
of $\, \mathcal{I}_\mathcal{G} \,$ are computed by
considering any set of
$\,|\mathcal{G}|- c(\mathcal{G})+ 1\,$ of the
unknowns $w_g$.  There exists a linear combination
of these $w_g$  which vanishes 
on $\mathcal{L}_{\mathcal{G}}$. If this linear
combination is unique (up to scaling) then
it is a circuit. The converse holds as well:
all circuits in $\, \mathcal{I}_\mathcal{G} \,$ 
are found in this manner.

\begin{ex}
\label{moregold} \rm
Let $\mathcal{G}$ be the genotype space in
Example \ref{threesix} and Figure \ref{fig:gold}.
We have $|\mathcal{G}| = 6$ and
$ c(\mathcal{G}) = 2$, so our bounds say that the number of
circuits is between two and six. In fact, this example
has precisely four circuits:
\begin{eqnarray*}
  f & = & w_{100}-w_{101}-w_{110}+w_{111}, \\
  g & = & w_{000}-w_{011}-w_{100}+w_{111}, \\
  n & = &  w_{011}+w_{101}+w_{110}-w_{000}-2w_{111}, \\
  s & = &  w_{000}+w_{101}+w_{110}-w_{011}-2w_{100} .
\end{eqnarray*}
The names $f$, $g$, $n$, and $s$ were chosen to be consistent  with
the discussion of the ambient 3-cube in Example~\ref{circuit3}.
The signs of these four circuits
characterize the possible interactions
in any fitness landscape on the genotyope space
$\mathcal{G}$. Only eight of the $16$ possible
sign patters can occur, since
the linear forms $f$, $g$, $n$, and $s$ lie in the two-dimensional 
space  $\, \mathcal{I}_\mathcal{G} $.
See also Figure~\ref{fig:secgold}.
\qed
\end{ex}

For certain genotype spaces $\mathcal{G}$,
the interaction space $ \mathcal{I}_\mathcal{G}$
will have a distinguished basis consisting of
 {\em interaction coordinates}. The choice
of interaction coordinates depends  on
the genotope $\Pi_\mathcal{G}$ and on the
situation of interest. A natural
choice exists in the case of the $n$-cube,
where $l=2$ and $\mathcal{G} = \{0,1\}^n$.
Consider any binary string
$i_1 i_2 \cdots i_n$ which has at least
two entries that are $1$. For such a string 
$i_1 i_2 \cdots i_n$ we 
introduce the following element of $\mathcal{I}_\mathcal{G}$:
$$ u_{i_1 i_2 \cdots i_n}
\quad := \quad
\frac{1}{2^{n-1}} \cdot
\sum_{j_1=0}^1
\sum_{j_2=0}^1 \cdots
\sum_{j_n=0}^1
(-1)^{i_1 j_1 + i_2 j_2 + \cdots + i_n j_n }
\cdot w_{j_1 j_2 \cdots j_n}. $$
The number of these linear forms equals
$c(\mathcal{G}) = 2^n-n-1$, and they 
form a basis of $\mathcal{I}_\mathcal{G}$.
We call the $u_{i_1 i_2 \cdots i_n}$ the interaction
coordinates for the $n$-cube.

The linear transformation above
is the {\em Fourier transform} for the
group $\,(\mathbb{Z}_2 )^n$. It has appeared frequently  in the 
mathematical biology literature.
For instance, Feldman et al.\ (1974) and Karlin and Feldman (1970) 
used it to study equilibria of dynamical evolutionary systems on two
and three binary loci. It appears in linkage analysis 
(Hallgr\'{\i}msd\'ottir, 2005), and in phylogenetics, 
where it gives rise to Hadamard
conjugation (Hendy and Charleston, 1993).

\begin{ex} 
\label{circuit3} 
{\bf (The 3-cube) }
\rm We discuss the interaction space for 
the three-dimensional cube $(n=3, l=2)$.
The four interaction coordinates are
\begin{eqnarray*}
u_{110} \quad = &
(w_{000} + w_{001} ) +  (w_{110} + w_{111} ) -
(w_{010} + w_{011} ) -  (w_{100} + w_{101} ) \\
u_{101} \quad = &
(w_{000} + w_{010} ) +  (w_{101} + w_{111} ) -
(w_{001} + w_{011} ) -  (w_{100} + w_{110} ) \\
u_{011} \quad = &
(w_{000} + w_{100} ) +  (w_{011} + w_{111} ) -
(w_{001} + w_{101} ) -  (w_{010} + w_{110} ) \\
u_{111} \quad = & \,\,\,
(w_{000} + w_{011} + w_{101} + w_{110})
\,\, - \,\, (w_{100}+w_{010}+w_{001}+w_{111}).
\end{eqnarray*}
These four linear forms form a natural basis for
the interaction space $\mathcal{I}_{\{0,1\}^3}$.
The interaction coordinate $u_{110}$
measures the {\em marginal epistasis}
between locus $1$ and locus $2$,
and similarly for $u_{101}$ and $u_{011}$.
The last interaction coordinate
$u_{111}$ measures the {\em three-way interaction}
among the loci.

The $3$-cube has  twenty circuits
in three symmetry classes. We abbreviate the circuits of
the $3$-cube by the first twenty letters of the alphabet.
The first six
circuits corresponds to the six faces of the $3$-cube,
and they measure the {\em conditional epistasis}
between two loci when the allele at the third locus is fixed:
$$
\begin{matrix}
a & := & u_{110}+u_{111} & = & w_{000}-w_{010}-w_{100}+w_{110} \\
b & := & u_{110}-u_{111} & = & w_{001}-w_{011}-w_{101}+w_{111} \\
c & := & u_{101}+u_{111} & = & w_{000}-w_{001}-w_{100}+w_{101} \\
d & := & u_{101}-u_{111} & = & w_{010}-w_{011}-w_{110}+w_{111} \\
e & := & u_{011}+u_{111} & = & w_{000}-w_{001}-w_{010}+w_{011}\\
 f & := & u_{011}-u_{111} & = & w_{100}-w_{101}-w_{110}+w_{111}.
\end{matrix}
$$
The second class of circuits relates
marginal epistases of two pairs of loci:
$$
\begin{matrix}
g & := & u_{110}+u_{101} & = & w_{000}-w_{011}-w_{100}+w_{111} \\
h & := & u_{110}-u_{101} & = & w_{001}-w_{010}-w_{101}+w_{110}   \\
i  & := & u_{110}+u_{011} & = & w_{000}-w_{010}-w_{101}+w_{111}  \\
j  & := & u_{110}-u_{011} & = & w_{001}-w_{011}-w_{100}+w_{110}    \\
k & := & u_{101}+u_{011} & = & w_{000}-w_{001}-w_{110}+w_{111}   \\
l  & := & u_{101}-u_{011} & = & w_{010}-w_{011}-w_{100}+w_{101}   .
\end{matrix}
$$
Geometrically, the six circuits
$g$, $h$, $i$, $j$, $k$, and $l$ correspond to squares formed by vertices
of the $3$-cube that slice the $3$-cube into two triangular prisms.

The last class consists of eight circuits which
relate the three-way interaction to the total two-way epistasis,
where signs are taken into consideration:
$$
\begin{matrix}
m & := &
-u_{011}-u_{101}-u_{110}-u_{111} \! & = \! & w_{001}+w_{010}+w_{100}-w_{111}-2w_{000} \\
n & := & 
-u_{011}-u_{101}-u_{110}+u_{111} \! & = \! & w_{011}+w_{101}+w_{110}-w_{000}-2w_{111} \\
o  & := & 
u_{011}+u_{101}-u_{110}-u_{111} \! & = \! & w_{010}+w_{100}+w_{111}-w_{001}-2w_{110} \\
 p & := & 
u_{011}+u_{101}-u_{110}+u_{111} \! & = \! & w_{000}+w_{011}+w_{101}-w_{110}-2w_{001} \\
 q & := & 
u_{011}-u_{101}+u_{110}-u_{111} \! & = \! & w_{001}+w_{100}+w_{111}-w_{010}-2w_{101} \\
r  & := & 
u_{011}-u_{101}+u_{110}+u_{111} \!  & =\! & w_{000}+w_{011}+w_{110}-w_{101}-2w_{010} \\
s & := & 
-u_{011}+u_{101}+u_{110}+u_{111} \! & = \! & w_{000}+w_{101}+w_{110}-w_{011}-2w_{100} \\
t & := & 
-u_{011}+u_{101}+u_{110}-u_{111} \! & = \! & w_{001}+w_{010}+w_{111}-w_{100}-2w_{011} .
\end{matrix}
$$
Geometrically, these correspond to the eight bipyramids in the $3$-cube. \qed
\end{ex}

The sign of the interaction coordinate $u$ in the two locus case 
of Example~\ref{twotriangles} determines the nature of the epistatic interaction.
In the three-locus case, one
may wish to record the  signs of each of the twenty 
circuits $a$, $b$, $\ldots$, $t$.
For instance, the signs of $a$, $b$, $c$, $d$, $e$, and $f$
specify all the conditional epistases of the fitness landscape $w$. 
The signs of the bipyramidal circuits $m$, $n$, $\ldots$, $t$
describe a three-way interaction which does not
have a two-locus interpretation.
The complete list of the twenty signs 
characterizes all the interactions 
among the genotypes, and also determines (but is not equivalent to) 
the shape of the fitness landscape. 
This is made precise in Proposition~\ref{prop:circuitpattern}.

While the detailed study of the $n$-cube for small values of $n$
provides a useful tool for data analysis, we wish to recall
that biological genotopes may not be cubes.
In such cases, there may not be any
choice of interaction coordinates 
which is as nice and canonical as that coming from the
Fourier transform. Fixing a basis for 
the interaction space will be a matter of choice
and preference. What remains canonical and natural 
is the full collection of all circuits of the genotope.

We conclude our discussion of gene interactions
by comparing the proposed circuits to the more traditional
approach of using ANOVA (Lindman, 1974). 
We illustrate the key difference between the two methods
in the following example. 

\begin{ex}   \label{2DNA} 
{\bf (Two DNA loci)} \rm 
Consider the genotype space $\mathcal{G}$ = 
\{\texttt{A}, \texttt{C}, \texttt{G}, \texttt{T}\}$^2$
of two DNA loci. A fitness landscape on $\mathcal{G}$
is a matrix
\[
   w = \left( \begin{array}{cccc}
     w_{\tt AA} &  w_{\tt AC} &  w_{\tt AG} &  w_{\tt AT} \\ 
     w_{\tt CA} &  w_{\tt CC} &  w_{\tt CG} &  w_{\tt CT} \\ 
     w_{\tt GA} &  w_{\tt GC} &  w_{\tt GG} &  w_{\tt GT} \\ 
     w_{\tt TA} &  w_{\tt TC} &  w_{\tt TG} &  w_{\tt TT}
   \end{array} \right). 
\]
Let $\bar{w}_{i \bullet}$ and $\bar{w}_{\bullet j}$ denote the row
and column means, respectively, of $w$, and denote by
$\bar{w}_{\bullet \bullet}$ the grand mean. In the 2-way ANOVA analysis
of the table $w$, one considers the 16 linear forms
\[
   w_{ij} - \bar{w}_{i \bullet} - \bar{w}_{\bullet j} 
     + \bar{w}_{\bullet \bullet} \qquad 
     (i,j \in \{{\tt A},{\tt C},{\tt G},{\tt T}\}),
\]
which measure the direction and amount by which each fitness value differs
from our expectation based only on the row and column means. 
By contrast, there are 204 circuits in the interaction space 
$\mathcal{I}_\mathcal{G}$, including, for example,
\begin{eqnarray*}
&   w_{\tt AG} - w_{\tt AT} - w_{\tt TG} + w_{\tt TT}, \\
& - w_{\tt AC} + w_{\tt AG} + w_{\tt CA} - w_{\tt CT} 
  - w_{\tt GG} + w_{\tt GT} - w_{\tt TA} + w_{\tt TC}, \\
& - w_{\tt AA} + w_{\tt AG} - w_{\tt GG} + w_{\tt GT} + w_{\tt TA} - w_{\tt TT}.
\end{eqnarray*}
Thus, the circuits measure deviation from linearity in a different, much finer
way than ANOVA does. \qed
\end{ex}

%%%%%%%%%%%%%%%%%%%%%%%%%%%%%%%%%%%%%%%%%%%%%%%%%%%%%%%%%%%%%%%%%%%%%%%%%
\bigskip
\setcounter{chapter}{4}
\setcounter{equation}{0} %-1
\noindent {\bf 4. The Shapes of Fitness Landscapes}
\smallskip

We have defined fitness landscapes as discrete objects
that assign
one fitness value to each individual genotype $g \in \mathcal{G}$.
However, in order to speak about ``shape'' or ``curvature''
of $\,w : \mathbb{R} \rightarrow \mathcal{G}$,
one needs a continuous object. 
This dilemma is resolved by  considering populations
$p \in \Delta_\mathcal{G}$ rather than individuals.

The fitness of a population $p$ is defined as the average fitness
of all individuals in the population. Since the individuals
are grouped into classes of identical genotypes, the population
fitness can be written as the dot product
\[
   w \cdot p \quad = \quad \sum_{g \in \mathcal{G}}
               w_g \cdot p_g.
\]
This notion of population fitness leads to an extension of a discrete
landscape $\,w :\mathcal{G} \rightarrow \mathbb{R}\,$ to
a function $\tilde{w}$ on the full genotope $\Pi_\mathcal{G}$.
The {\em continuous landscape} 
$\,\tilde{w} \colon \Pi_\mathcal{G} \rightarrow \mathbb{R}\,$
derived from $w$
assigns to each point $v$ in the genotope the maximum fitness among all
populations $p$ with these allele frequencies. We define
\begin{equation}   \label{LP}
  \tilde{w}(v) \,\,\,\, :=\, \,\,\,\max \, \left\{ p \cdot w \,:\,
  p \in \rho^{-1}(v) \right\} \qquad
  \mbox{ for all } v \in \Pi_{\mathcal{G}}.  
\end{equation}
Computing the fitness value $\tilde{w}(v)$ 
amounts to solving a $c({\mathcal{G}})$-dimensional linear 
programming problem, namely, to maximizing the linear 
functional $\,p \mapsto p \cdot w\,$ over the fiber $\rho^{-1}(v)$.
The continuous landscape $\tilde{w}$
is the smallest convex function which has the 
same values as $w$ on the vertices of $\Pi_\mathcal{G}$.  
It is continuous and piecewise linear.
Our classification of landscapes rests on the following remark.

\begin{rmk}
\label{rmk:domains}
The domains of linearity of the piecewise linear function $\, \tilde{w} \,$
are the cells in a regular polyhedral subdivision $\, \Pi_\mathcal{G}[w]\,$
of the genotope $\Pi_\mathcal{G}$.
\end{rmk}

We refer to the book of De~Loera et al.\ (2006)
for an introduction to the geometry of polyhedral subdivisions.
Remark \ref{rmk:domains} appears in
(De~Loera et al., 2006, Chap.~2).
  We call the induced polyhedral subdivision 
 $\Pi_\mathcal{G}[w]$ the {\em shape} of
the fitness landscape~$w$.
For almost all fitness landscapes $w \in \mathbb{R}^\mathcal{G}$,
the subdivision $\Pi_{\mathcal{G}}[w]$ will be a {\em regular triangulation}, 
i.e., a subdivision all of whose cells are simplices. 
We call such landscapes {\em generic fitness landscapes}.

The simplices in the triangulation $\Pi_{\mathcal{G}}[w]$
have a natural interpretation in terms of populations.
For any $n$-tuple of allele frequencies $v \in \Pi_{\mathcal{G}}$, 
there is a  unique fittest population $p$  with $\rho(p) = v$.
The genotypes that occur in this fittest population
are the vertices of the simplex of $\Pi_{\mathcal{G}}[w]$
which contains $v$. Thus, knowing the shape of 
a fitness landscape $w$ is  equivalent to knowing 
all the fittest populations for $w$.
For instance, in Example~\ref{twotriangles},
if $w$ has positive epistasis, then 
$01$ and $10$ cannot coexist in a fittest population,
so any fittest population consists either of genotypes in
the triangle $\{00,01,11\}$ or of genotypes in $\{00,10,11\}$.

The number of shapes of fitness landscapes on
a fixed genotype space $\mathcal{G}$ is finite.
If $\mathcal{G}$ has few elements (say, less than twenty), 
then a complete list of all generic shapes can be compiled
using the software {\tt TOPCOM} (Rambau, 2002)
which enumerates triangulations.
 For instance,
the number of generic shapes of fitness landscapes
on the cube $\{0,1\}^n$ is two if $n=2$,
and it is $74$ if $n=3$ (Table~\ref{tab:74}).
In Section~5, we discuss the $74$ shapes
of fitness landscapes on $\{0,1\}^3$. 

\smallskip

In order to classify all shapes of fitness landscapes 
on a genotype space $\mathcal{G}$, we need to list all
polyhedral subdivisions of the genotope $\Pi_{\mathcal{G}}$.
This set of subdivisions is represented by the
{\em secondary polytope} $\Sigma_{\mathcal{G}}$
(De~Loera et al., 2006, Chap.~5).
The secondary polytope $\Sigma_\mathcal{G}$
has dimension $c(\mathcal{G})$, it
lives in the space dual to $\mathbb{R}^{\mathcal{G}}$,
and its vertices  are in bijection with the 
generic shapes. The higher-dimensional faces
of the secondary polytope  $\Sigma_\mathcal{G}$
correspond to non-generic shapes.
Thus, the secondary polytope of a genotype space $\mathcal{G}$
represents all shapes of fitness landscapes on $\mathcal{G}$
and their neighborhood relations.
In Example \ref{twotriangles},
the secondary polytope $\Sigma_\mathcal{G}$
is a line segment. Its two vertices
correspond to the two generic shapes, 
and the segment itself corresponds to the flat shape.

In general, the secondary polytope can be defined as follows.
Consider the average fitness of the fittest populations
over all $n$-tuples of allele frequencies:
$$ {\bf F}(w) \quad = \quad
\frac{1}{{\rm vol}(\Pi_\mathcal{G})} \cdot
\int_{\Pi_\mathcal{G}} \tilde{w}(v) \, dv. $$
This integral is a piecewise linear function in
the unknown fitness landscape $w \in \mathbb{R}^\mathcal{G}$. Two 
landscapes $w$ and $w'$ have the same generic shape precisely
when they lie in a common domain of linearity of
this piecewise-linear function. Thus for each
generic shape, the function ${\bf F}$ is represented
by a linear functional on $\mathbb{R}^\mathcal{G}$,
i.e., by a vector $F$
in the dual space  $(\mathbb{R}^\mathcal{G})^*$.
The coordinate $F_g$ of this vector
with respect to the standard basis on  $(\mathbb{R}^\mathcal{G})^*$,
equals the probability
that the genotype $g$ appears in a fittest population.
The secondary polytope $\Sigma_\mathcal{G}$ is the
convex hull in  $(\mathbb{R}^\mathcal{G})^*$ of these
vectors $F$, one for each generic shape.

\begin{ex} \label{gold} \rm
Let $\, \mathcal{G}  =  \{ 
 000, \, \, 011,\,  100, \, 101,\, 110,  \, 111 \} \,$ be the genotype space  in Example \ref{threesix}
and Figure \ref{fig:gold}. The two interaction coordinates are
\begin{eqnarray*}
 x & = & w_{100} + w_{111} - w_{101} - w_{110} \\
 y & = & w_{000} + w_{111} - w_{011} - w_{100}.
\end{eqnarray*}
The average fitness ${\bf F}(w)$
is the piecewise-linear
function in the six fitness values
$w_{000}$, $w_{011}$, $w_{100}$, $w_{101}$, $w_{110}$, and $w_{111}$.
The function ${\bf F}(w)$ equals
$$
\frac{1}{4} \cdot {\rm max} \left\{
\begin{matrix}
2 w_{000} + 4 w_{011} +  4 w_{100} + 2 w_{101} + 2 w_{110} + 2 w_{111} , \\
2 w_{000} + 4 w_{011} +  3 w_{100} + 3 w_{101} + 3 w_{110} +   w_{111} , \\
3 w_{000} + 3 w_{011} +    w_{100} + 4 w_{101} + 4 w_{110} +   w_{111} , \\
4 w_{000} + 2 w_{011} +    w_{100} + 3 w_{101} + 3 w_{110} + 3 w_{111} , \\
4 w_{000} + 2 w_{011} +  2 w_{100} + 2 w_{101} + 2 w_{110} + 4 w_{111} 
\end{matrix}
\right\}.
$$
Using the interaction coordinates, the average fitness 
can be rewritten as follows:
\begin{eqnarray*}
 {\bf F}(w) \quad = 
& \!\!\!\!\ (1/4) \cdot 
{\rm max} \{\, 0, \,  x, \, 2x-y, \, x-2y,\, -2y \, \}
\\ & \quad + \,
 w_{011} +  w_{100} + 
( w_{000}  +  w_{101} +  w_{110} +  w_{111} ) /2.
\end{eqnarray*}
The five cases in this maximum correspond to the
five possible shapes of the fitness landscape.
The corresponding triangulations of $\Pi_\mathcal{G}$ are
{\small
\begin{eqnarray*}
\bigl\{ \{101, 111, 100, 011\}, 
\{110, 111, 100, 011\},
\{101, 000, 100, 011\},  
\{110, 000, 100, 011\} \bigr\} \\
\bigl\{ \{101, 000, 100, 011\}, 
\{110, 000, 100, 011\}, 
\{110, 101, 100, 011\}, 
\{110, 101, 111, 011\}\bigr\}  \\
\bigl\{ \{110, 101, 111, 011\}, 
\{110, 101, 000, 100\},
\{110, 101, 000, 011\}_{\hbox{has volume $2$}}
\, \bigr\}   \qquad \quad
\\
\bigl\{\{110, 101, 000, 100\}, 
\{110, 101, 000, 111\}, 
\{101, 000, 111, 011\}, 
\{110, 000, 111, 011\}\bigr\}  \\
\bigl\{\{101, 000, 111, 100\}, 
\{110, 000, 111, 100\}, 
\{101, 000, 111, 011\}, 
\{110, 000, 111, 011\}\bigr\}.
\end{eqnarray*}
}
The secondary polytope 
$\Sigma_\mathcal{G}$ is a pentagon
whose vertices are labeled
by the five triangulations.
It is represented geometrically
as the pentagon in the interaction
space $\mathcal{I}_\mathcal{G}$ 
whose directed edges are
$x$, $x-y$, $-x-y$, $-x$, and $2y$.
See Figure~\ref{fig:secgold}
for an illustration of the pentagon $\Sigma_\mathcal{G}$ 
in the context of HIV fitness data.
\qed
\end{ex}

We now explain the relationship between the
shapes of fitness landscapes on $\mathcal{G}$
and the circuits in the interaction space 
$\mathcal{I}_\mathcal{G}$.
We define the {\em circuit sign pattern} of
a fitness landscape $w \in \mathbb{R}^{\mathcal{G}}$
to be the list which records  the
sign (positive, zero or negative)
of the numerical value of each circuit at $w$.

\begin{prop}   \label{prop:circuitpattern}
The shape $\Pi_\mathcal{G}[w]$ of a
fitness landscape $w \in \mathbb{R}^\mathcal{G}$ is determined by its circuit
pattern, but the converse generally does not hold.
\end{prop}

\begin{proof}
Both the circuit sign patterns and the shapes define
a subdivision of $\mathbb{R}^\mathcal{G}$ into cones
which fit together to form a fan. This ensures that
we need only consider the generic case when all circuits
are either positive or negative at $w$.
Each such linear inequality can be written in the form
$$ 
\sum_{g \in \mathcal{G}_1} \alpha_g w_g 
\quad > \quad 
\sum_{g \in \mathcal{G}_2} \beta_g w_g,
$$
where 
 $\mathcal{G}_1$ and   $\mathcal{G}_2$ are
disjoint subsets of  $\mathcal{G}$, and the 
$\alpha_g$ and $\beta_g$ are positive reals.
The genotypes in $\mathcal{G}_2$
cannot coexist in a fittest population, since
they can be replaced by the genotypes in $\mathcal{G}_1$,
thus increasing  population fitness while
keeping the allele frequencies unchanged.
Containing such a replaceable set $\mathcal{G}_1$
is the only obstruction to being a simplex in the
triangulation $\Pi_\mathcal{G}[w]$. In other words,
the sets $\mathcal{G}_1$ derived from circuits
as above are the minimal non-faces
of the triangulation $\Pi_\mathcal{G}[w]$.
This proves that $\Pi_\mathcal{G}[w]$ is
determined by the circuit sign patterns at $w$.

The second part of the proposition follows from
Figure~\ref{fig:secgold}, There
are precisely four circuits, namely,
$x$, $y$, $x+y$, $x-y$, and hence eight possible
$\{+/-\}$-sign patterns. The eight sign patters
map correspond to only five distinct shapes.
\end{proof}
\par

%%%%%%%%%%%%%%%%%%%%%%%%%%%%%%%%%%%%%%%%%%%%%%%%%%%%%%%%%%%%%%%%%%%%%%%%%
\bigskip
\setcounter{chapter}{5}
\setcounter{equation}{0} %-1
\noindent {\bf 5. Three-way Interactions}
\smallskip

We illustrate our generalization of the classical two-locus two-allele
situation (Figure \ref{fig:updown}) by examining the possible fitness shapes for three biallelic loci. 
Here, the genotype space is 
 $\mathcal{G} = \{000, 001, 010, 011, 100, 101, 110, 111 \}$,
and the possible fitness shapes are exactly the 74 triangulations of the 
3-cube (Table~\ref{tab:74}). 

Each triangulation is uniquely represented
by its  {\em GKZ vector} (De~Loera et al., 2006),
which is the vector $F \in  (\mathbb{R}^\mathcal{G})^*$
introduced prior to Example~\ref{gold}. The GKZ vector indicates for
each vertex $g \in \mathcal{G}$ the sum of the normalized volumes
of all tetrahedra in the given triangulation that contain $g$. 
Equivalently, if the allele frequency vector is chosen uniformly
at random from  $\Pi_{\mathcal{G}}$, then
the $g$-th entry of the GKZ vector is 
the probability that genotype $g$ appears in the fittest population.
We refer to the shape of a fitness landscape for a three-locus two-allele 
system by its number (1 to 74) as appearing in the first column 
of Table~\ref{tab:74}.

\begin{table}
\centering
\begin{tabular}{rcc|rcc}
\#/T & GKZ & Out-edges  
   & \#/T & GKZ & Out-edges  \\\hline
 1/1 & 15515115 & 3t4q5o6m  &   38/4 & 31355313 & 39$\overline{l}$44$\overline{g}$51c59d \\
 2/1 & 51151551 & 7s8r9p10n &   39/4 & 31533513 & 38l44$\overline{i}$53e60f \\
 3/2 & 14436114 & 1$\overline{t}$11b13d17$\overline{e}$&   40/4 & 33155133 & 42$\overline{j}$45$\overline{g}$54a61b \\
 4/2 & 14614314 & 1$\overline{q}$12b14f18$\overline{c}$ &   41/4 & 33511533 & 43$\overline{h}$46$\overline{i}$55a62b \\
 5/2 & 16414134 & 1$\overline{o}$15d16f19$\overline{a}$ &   42/4 & 35133153 & 40j45$\overline{k}$57e63f \\
 6/2 & 34414116 & 1$\overline{m}$28$\overline{e}$29$\overline{c}$31$\overline{a}$ &   43/4 & 35311353 & 41h46$\overline{k}$58c64d \\ 
 7/2 & 41163441 & 2$\overline{s}$20a22c26$\overline{f}$ &   44/4 & 51333315 & 38g39i65$\overline{b}$68$\overline{a}$ \\
 8/2 & 41341641 & 2$\overline{r}$21a23e27$\overline{d}$ &   45/4 & 53133135 & 40g42k66$\overline{d}$69$\overline{c}$ \\
 9/2 & 43141461 & 2$\overline{p}$24c25e30$\overline{b}$ &   46/4 & 53311335 & 41i43k67$\overline{f}$70$\overline{e}$ \\
10/2 & 61141443 & 2$\overline{n}$32$\overline{f}$33$\overline{d}$34$\overline{b}$ &   47/5 & 13356222 & 11$\overline{d}$13$\overline{b}$35f71$\overline{e}$ \\
11/3 & 13446213 & 3$\overline{b}$12$\overline{l}$47d51$\overline{e}$ &   48/5 & 13623522 & 12$\overline{f}$14$\overline{b}$36d72$\overline{c}$ \\
12/3 & 13624413 & 4$\overline{b}$11l48f53$\overline{c}$ &   49/5 & 16323252 & 15$\overline{f}$16$\overline{d}$37b73$\overline{a}$ \\
13/3 & 14346123 & 3$\overline{d}$15$\overline{j}$47b54$\overline{e}$ &   50/5 & 22265331 & 20$\overline{c}$22$\overline{a}$35e71$\overline{f}$\\
14/3 & 14613423 & 4$\overline{f}$16$\overline{h}$48b55$\overline{c}$ &   51/5 & 22356213 & 11e17$\overline{b}$38$\overline{c}$71d \\
15/3 & 16324143 & 5$\overline{d}$13j49f57$\overline{a}$ &   52/5 & 22532631 & 21$\overline{e}$23$\overline{a}$36c72$\overline{d}$ \\
16/3 & 16413243 & 5$\overline{f}$14h49d58$\overline{a}$ &   53/5 & 22623513 & 12c18$\overline{b}$39$\overline{e}$72f\\
17/3 & 23346114 & 3e28$\overline{g}$51b54d &   54/5 & 23256123 & 13e17$\overline{d}$40$\overline{a}$71b \\
18/3 & 23613414 & 4c29$\overline{i}$53b55f &   55/5 & 23612523 & 14c18$\overline{f}$41$\overline{a}$72b \\
19/3 & 26313144 & 5a31$\overline{k}$57d58f &   56/5 & 25232361 & 24$\overline{e}$25$\overline{c}$37a73$\overline{b}$ \\
20/3 & 31264431 & 7$\overline{a}$21$\overline{l}$50c59$\overline{f}$ &   57/5 & 26223153 & 15a19$\overline{d}$43$\overline{e}$73f \\
21/3 & 31442631 & 8$\overline{a}$20l52e60$\overline{d}$ &   58/5 & 26312253 & 16a19$\overline{f}$43$\overline{c}$73d \\
22/3 & 32164341 & 7$\overline{c}$24$\overline{j}$50a61$\overline{f}$ &   59/5 & 31265322 & 20f26$\overline{a}$38$\overline{d}$71c \\
23/3 & 32431641 & 8$\overline{e}$25$\overline{h}$52a62$\overline{d}$ &   60/5 & 31532622 & 21d27$\overline{a}$39$\overline{f}$72e \\
24/3 & 34142361 & 9$\overline{c}$22j56e63$\overline{b}$ &   61/5 & 32165232 & 22f26$\overline{c}$40$\overline{b}$71a \\
25/3 & 34231461 & 9$\overline{e}$23h56c64$\overline{b}$ &   62/5 & 32521632 & 23d27$\overline{e}$41$\overline{b}$72a \\
26/3 & 41164332 & 7f32$\overline{g}$59a61c &   63/5 & 35132262 & 24b30$\overline{c}$42$\overline{f}$73e \\
27/3 & 41431632 & 8d33$\overline{i}$60a62e &   64/5 & 35221362 & 25b30$\overline{e}$32$\overline{d}$73c \\
28/3 & 43324116 & 6e17g65$\overline{c}$66$\overline{a}$ &   65/5 & 52323216 & 28c29e44b74$\overline{a}$ \\
29/3 & 43413216 & 6c18i65$\overline{e}$67$\overline{a}$ &   66/5 & 53223126 & 28a31e45d74$\overline{c}$ \\
30/3 & 44131362 & 9b34$\overline{k}$63c64e &   67/5 & 53312226 & 29a31c46f74$\overline{e}$ \\
31/3 & 44313126 & 6a19k66$\overline{e}$67$\overline{c}$ &   68/5 & 61232325 & 32d33f44a74$\overline{b}$ \\
32/3 & 61142334 & 10f26g68$\overline{d}$69$\overline{b}$ &   69/5 & 62132235 & 32b34f45c74$\overline{d}$ \\
33/3 & 61231434 & 10d27i68$\overline{f}$70$\overline{b}$ &   70/5 & 62221335 & 33b34d46e74$\overline{f}$ \\
34/3 & 62131344 & 10b30k69$\overline{f}$70$\overline{d}$ &   71/6  & 22266222 & 47e50f51$\overline{d}$54$\overline{b}$59$\overline{c}$61$\overline{a}$ \\
35/4 & 13355331 & 36$\overline{l}$37$\overline{j}$47$\overline{f}$50$\overline{e}$ &   72/6 & 22622622 & 48c52d53$\overline{f}$55$\overline{b}$60$\overline{e}$62$\overline{a}$ \\
36/4 & 13533531 & 35l37$\overline{h}$48$\overline{d}$52$\overline{c}$ &   73/6 & 26222262 & 49a56b57$\overline{f}$58$\overline{d}$63$\overline{e}$64$\overline{c}$ \\
37/4 & 15333351 & 35j36h49$\overline{b}$56$\overline{a}$ &  74/6 & 62222226 & 65a66c67e68b69d70f \\
\end{tabular}
\caption{The 74 shapes of fitness landscapes of the three-locus two-allele system.
The first column specifies the shape number and type, the second
column the GKZ vector, and the third column
the out-edges in the secondary polytope.
\label{tab:74}
}
\end{table}

The $74 $ shapes label the vertices of the secondary polytope 
$\Sigma_\mathcal{G}$ of the 3-cube
and are listed in  Table~\ref{tab:74}. 
As an example consider the 17th vertex of $\Sigma_\mathcal{G}$:
\begin{center} 
\begin{tabular}{rcc}
17/3 & 23346114 & 3e28$\overline{g}$51b54d.
\end{tabular}
\end{center}
\noindent
This says that shape~17 has the GKZ vector $(2,3,3,4,6,1,1,4)$. The third column says
 that shape~17 is adjacent to the shapes~3, 28, 51 and 54. The letters
 $a$, $b$, $\ldots$, $t$ refer to the list of circuits in Example \ref{circuit3}.
 The GKZ vector $(2,3,3,4,6,1,1,4)$  differs  from the
 GKZ vector of  shape~3 by the circuit
 $$e \,\,\,\, = \,\,\,\,  u_{011}+u_{111} \,\,=\,\,
 w_{000}-w_{001}-w_{010}+w_{011}
 \,\,\,\, = \,\,\,\,(1,-1,-1,1,0,0,0,0). $$
  Similarly, shape~17
differs from shape~28 by the circuit $-g = -u_{110}-u_{101}$, from shape 51 
by the circuit $b = u_{110}-u_{111}$, 
and from shape~54 by the circuit $d = u_{101}-u_{111}$. 
The circuits $e$, $b$, and $d$ measure conditional epistasis between 
two loci when the third is fixed, so the secondary polytope 
tells us, for example, that shapes~17 and 51 differ 
only by such a conditional epistasis interaction. 

In the first column of Table~\ref{tab:74}, one more number of interest 
is displayed after the shape number and separated by a slash, namely
the {\em interaction type}.
Shapes of the same interaction type differ only in the labeling of the 
vertices of the cube; in other words, they correspond
to the same combinatorial type of triangulation. 
The six interaction types correspond to the six symmetry
classes of triangulations of the 3-cube. These six 
classes are depicted in (De~Loera et al., 2006, Fig.~1.38) and in
(Grier et al., 2006, Fig.~1), and we refer to these references 
for further mathematical background and generalizations
to higher dimensions.

Table~\ref{tab:74} specifies all the shapes of fitness 
landscapes for three biallelic loci. It is useful to examine
the six interaction types, and to consider their 
biological meaning.
The two triangulations of type~1 
are obtained if the three-way interaction $u_{111}$ is either
very large or very small. It
consists of a central tetrahedron
of volume two and four adjacent tetrahedra of volume one. 
The central tetrahedron can either use
the genotypes with an even number of mutations, or those with an
odd mutation count. 
Fitness landscapes of this type are linear on the central tetrahedron
and on the adjacent tetrahedra. The curvature is such that fitness
decreases more strongly than expected towards any of the genotypes
that are not part of the central tetrahedron. These four
genotypes are ``sliced off'', and the
fittest populations avoid these genotypes whenever possible.
Thus, the shape being of type~1 means that the
fittest populations include either 
all genotypes with an even number of mutations,
or all genotypes with an odd number of mutations. 

Type~6 landscapes can be regarded as opposite to
type~1. The four triangulations of this type are 
indexed by the four different diagonals of the 3-cube. 
Each diagonal induces a triangulation that
consists of six tetrahedra arranged around the diagonal in such a way
that each tetrahedron has exactly two adjacent tetrahedra. Fitness
landscapes of this type are linear on each tetrahedron and the curvature
is such that any of the six tetrahedra has a higher fitness than expected
from its two neighbors. No single genotypes are sliced off,
as all entries of the GKZ vector are bigger than $1$.
For example, shape~74 uses  the diagonal through
000 and 111. Hence the vertices of the tetrahedra correspond to different
accumulative mutational pathways of length four from 000 to 111. For
example, $000 \to 001 \to 101 \to 111$ is one such pathway. 
Thus, the fittest populations involve all genotypes from exactly one
of the six possible pathways.

Types~1 and 6 represent the two extreme 
types of three-way interaction. Type 1 emphasizes the number of mutations 
irrespective of the particular context or pathway they occur in.
By contrast, type~6 emphasizes the mutational pathway.
Type~1 fitness shapes occur when intermediate types
are fitter than expected irrespective of the specific intermediate genotype. 
Likewise, we expect type~6 fitness shapes whenever higher
fitness values than expected occur only in specific genotypic contexts. 
The remaining interaction types (2 to 5) are intermediate and 
share features from both type~1 and type~6.
%For example, the 24
%triangulations of type~5 are indexed by a diagonal and one 
%other vertex. That vertex is sliced off, and the remaining
%polytope is divided into a pentagonal ring of tetrahedra around
%the diagonal. Similarly to type~6, the curvature is such that
%any of the five tetrahedra has a higher fitness than expected
%from its two neighbors. 

It is important to note that the third column in 
Table~\ref{tab:74} gives the minimal
set of linear inequalities that characterize
the {\em robustness  cone} for each of the $74$ shapes.
These cones consist of all fitness landscapes
that have the given shape. 
For instance, shape 74 is characterized by 
the inequalities $a,b,c,d,e,f > 0$ which says that
conditional epistasis is positive for any fixation
of one of the three loci. While type 6 requires
six inequalities, each of the other five types
requires only four. For instance,
being in shape 38 means that
the conditional epistases $c$ and $d$ are positive
while the marginal epistases $g$ and $l$ are negative.

For the four-locus two-allele system ($\mathcal{G}  = \{0,1\}^4$),
it was shown by Grier et al.\ (2006) 
that the number of shapes is $87,959,448$,
and that these fall into $235,277$ symmetry types.
A table analogous to Table \ref{tab:74}, listing
one representative shape from each type, is available at
\url{http://bio.math.berkeley.edu/4cube/}.
\par

%%%%%%%%%%%%%%%%%%%%%%%%%%%%%%%%%%%%%%%%%%%%%%%%%%%%%%%%%%%%%%%%%%%%%%%%%
\bigskip
\setcounter{chapter}{6}
\setcounter{equation}{0} %-1
\noindent {\bf 6. Positive Epistasis in HIV?}
\smallskip

In this section, we characterize the fitness landscape of HIV-1.
We use the data published in (Segal et al., 2004), which is also
described in (Bonhoeffer et al., 2004). The data consists of 288
genotype--fitness pairs. The reported genotype is
the HIV protein sequence composed of positions 4 to 99 of the protease (PRO)
and positions 38 to 223 of the reverse transcriptase (RT). Fitness
was measured as the number of offspring in a single replication
cycle and was reported relative to the fixed wild type strain 
NL4-3 on a logarithmic scale.
Univariate and 
multivariate analyses show that mutation L90M in the protease, and mutations
M184V and T215Y in the RT are the major determinants of
fitness in this data set. The notation L90M means
that at position 90, the amino acid leucine (L) is replaced by
methionine (M).
We therefore analyze the fitness
shape of this three-locus two-allele system.
We identify the subsets of \{L90M,
M184V, T215Y\} with binary strings of length
three. For instance, the string 010 in the third row 
in Table~\ref{tab:hiv-ls}
is the genotype carrying only mutation M184V.
%, but neither L90M nor T215Y.

\begin{table}
\centering
\begin{tabular}{crrrrrrr}
Genotype & Count & Min. & 1st Qu. & Median & \bf Mean & 3rd Qu. & Max.\\\hline
000 & 214 & 0.1917 & 1.4770&  1.6410& \bf 1.5800& 1.7910&  2.0530\\ 
001 & 5 & 0.5344&  0.6990&  1.1880& \bf 1.1950& 1.7710&  1.7850\\ 
010 & 8 & --0.3355&  1.1440&  1.2960& \bf 1.1330 & 1.4870 & 1.5310\\ 
011 & 8 & 1.0000 &  1.2500&  1.5150& \bf 1.4300  & 1.6360 &  1.7240\\ 
100 & 7 & 0.4771&  1.3470&  1.4650& \bf 1.4410 & 1.7890 & 1.8750\\ 
101 & 13 & 0.3010&  0.8673&  1.3420& \bf 1.2320 & 1.5840 & 1.8870\\ 
110 & 11 & 0.6021&  1.1610&  1.3700& \bf 1.2940&  1.5370&  1.6920\\ 
111 & 22 & --0.4771&  0.9472&  1.1790 & \bf 1.0450 & 1.3850 & 1.7900\\\hline
\end{tabular}
\caption{HIV random fitness landscape on the three amino acid
mutations PRO L90M, RT M184V, and RT T215Y. The mean values 
(in bold) are used for significance testing.
\label{tab:hiv-ls}
}
\end{table}

The fitness values were measured
for different viruses, some of which share a mutational
pattern on the three selected loci. Thus, instead 
of a single fitness value for each genotype, we have a distribution.
A {\em random fitness landscape} on a genotype space $\mathcal{G}$
is a collection of $|\mathcal{G}|$ continuous random variables
$W = (W_g)_{g \in \mathcal{G}}$. We think of a realization 
of $W$ as a real-valued function
$w \colon \mathcal{G} \rightarrow \mathbb{R}$ on the
genotype space. 
Thus, $W$ takes values in $\mathbb{R}^{\mathcal{G}}$.
In general, stochastic fluctuations in $W$ can arise from
measurement noise in assessing the reproductive fitness
experimentally and from biological variation that is not
linked to the $n$ genetic loci.
The HIV random fitness landscape 
is summarized in Table~\ref{tab:hiv-ls}.

Following Bonhoeffer et al.\ (2004) and Sanju\'{a}n et al.\ (2004),
we first examine the total marginal two-way epistasis
\[
    E \,\, = \,\, u_{110} + u_{101} + u_{011},
\]
with  $u_{ijk}$ defined in Example \ref{circuit3}.
Computing $E$ means
 pooling epistasis estimates obtained from 
the three pairs of loci.
We use randomized fitness values to estimate $E$ under the
null hypothesis of no epistasis.
Unlike in (Bonhoeffer et al., 2004), we do not find
the observed positive value of $\hat{E} = 0.025$ to be 
significantly greater than zero ($P > 0.35$). 
This discrepancy may reflect the limited
statistical power of our analysis, which is based on a smaller
data set and on only 3 loci, although
Bonhoeffer et al.\ found positive epistasis to be more
pronounced when restricting to the most influential
sequence positions.

\begin{table}
\centering
\begin{tabular}{lccrr}
Circuit & Pair & Context & Cond.~epist. & $P$-value \\\hline
$a$ &  90--184 & T215 &   0.300 & 0.110  \\
$b$ &  90--184 & 215Y & --0.421 & 0.059 \\
$c$ &  90--215 & M184 &   0.175 & 0.230  \\
$d$ &  90--215 & 184V & --0.545 & 0.013 \\
$e$ & 184--215 & L90  &   0.682 & 0.008 \\
$f$ & 184--215 & 90M  & --0.039 & 0.410  \\\hline
\end{tabular}
\caption{Conditional 2-way epistasis in HIV. The $P$-value denotes
the fraction of epistasis values in the bootstrap 
sample that are lower (in the case of negative epistasis) or 
higher (in the case of positive epistasis) than the
observed mean conditional epistasis. Circuits are labeled as
in Example~\ref{circuit3}.
\label{tab:hiv2}
}
\end{table}

\begin{figure}
\centering
\includegraphics[angle=270,scale=0.4]{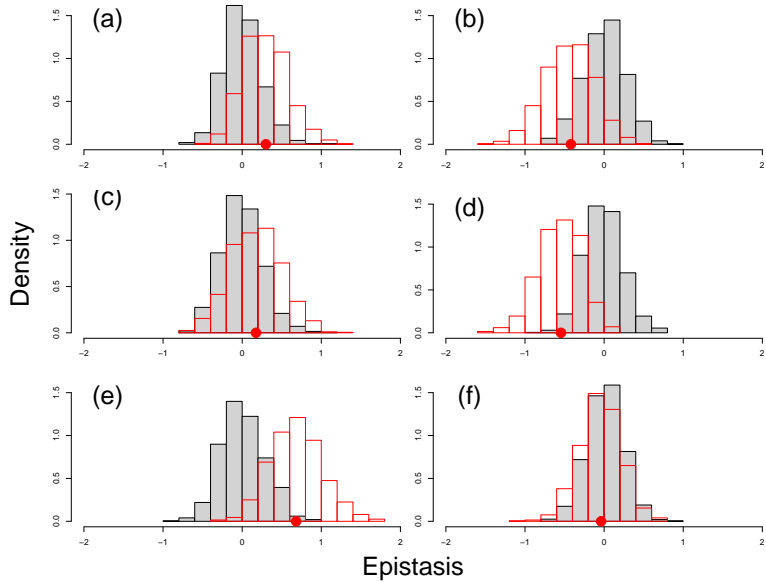}
\caption{Conditional two-way epistasis in HIV, grouped according
to circuits $a,\ldots,f$. }
\label{fig:hiv2}
\end{figure}

Rather than marginalizing, we next consider two-way epistasis 
conditional on the value at the third locus, i.e., 
we determine the pattern of epistasis
for each pair of loci in the context of the third locus
(Table~\ref{tab:hiv2}). For all three pairs, we find positive 
epistasis conditioned on the wild type allele at the third
position, and negative conditional epistasis otherwise. 
Two out of six comparisons reached statistical significance,
including one case of negative epistasis. 
Figure~\ref{fig:hiv2} shows the empirical null distributions
(grey histogram bars) and, in addition, the empirical 
distributions of the circuits $a$, $b$, $c$, $d$, $e$, and
$f$ (as defined in Example~\ref{circuit3}) that represent
conditional epistasis (clear histogram bars). 
The distributions of circuits are
obtained by resampling fitness values for the same genotype.
They reflect the uncertainty in the shape of the fitness
landscape due to multiple discordant measurements.
 We conclude that the HIV
fitness landscape on the three selected loci is 
insufficiently described as ``positively epistatic''.

\begin{figure}
\centering
\includegraphics[scale=.8]{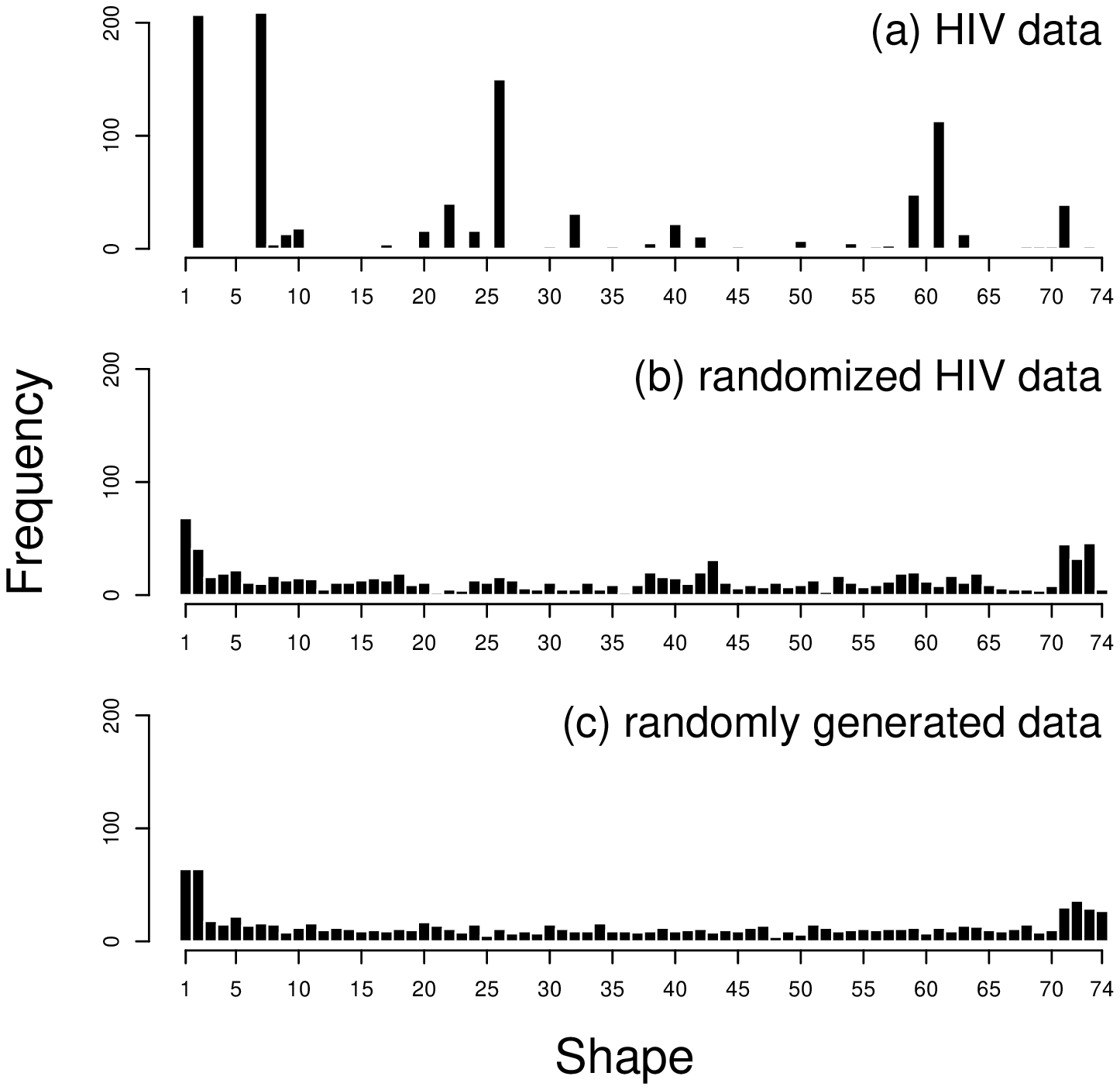}
\caption{Three-way epistasis in HIV, analyzed using the
74 shapes in Table \ref{tab:74}.
 Shape distribution
of (a) the HIV random fitness landscape on mutations PRO L90M, 
RT M184V, and RT T215Y; (b) the same fitness values 
randomly assigned to genotypes; and (c) a random landscape 
in which fitness values are identically and uniformly distributed.
}
\label{fig:hiv3}
\end{figure}

We now characterize the shape of the HIV fitness landscape by
estimating the distribution on the 74 
shapes in Table~\ref{tab:74} induced by the data 
in Table~\ref{tab:hiv-ls}.
We use a resampling 
scheme to estimate the distribution and shape of $W$.
 The distribution of $\Pi_\mathcal{G}[W]$ is
shown in Figure~\ref{fig:hiv3}(a).
The dominant shapes are \#7 (frequency 20.9\%),
\#2 (20.7\%), \#26 (15\%), and \#61 (11.3\%).
In Figure~\ref{fig:hiv3}(b),
we randomized the fitness values.
Comparing the two histograms shows that the observed
shape distribution is very different from a randomly
chosen landscape using the same values. The
histogram in Figure~\ref{fig:hiv3}(c) results from sampling
fitness values uniformly at random from $[0,1]$ and is
similar to the one in subfigure (b). Figure~\ref{fig:hiv3}
can be used to detect differences in fitness shape between random
landscapes, and to identify single shapes that appear with high probability.

The dominant shapes of the HIV fitness landscape are very similar
and have certain features in common.
For example, the GKZ vectors that correspond to shapes~2, 
7, 10, 26, and 32 all share a coordinate
$1$ in both the second and third position.
These five shapes account for 61.5\% of the probability mass.
They all slice off the two genotypes $001$ and $010$,
which correspond to the single mutants \{M184V\} and \{T215Y\},
respectively. Both mutations are known to 
reduce the fitness of HIV. Indeed, M184V
develops shortly after initiation of therapy with the 
antiviral drug lamivudine in most patients. Although
M184V carrying viruses are resistant to lamivudine, 
administration of the drug is often continued in order
to maintain the mutation and its fitness impairing effect
(Wainberg, 2004).

\begin{figure}
\centering
\includegraphics[scale=.65]{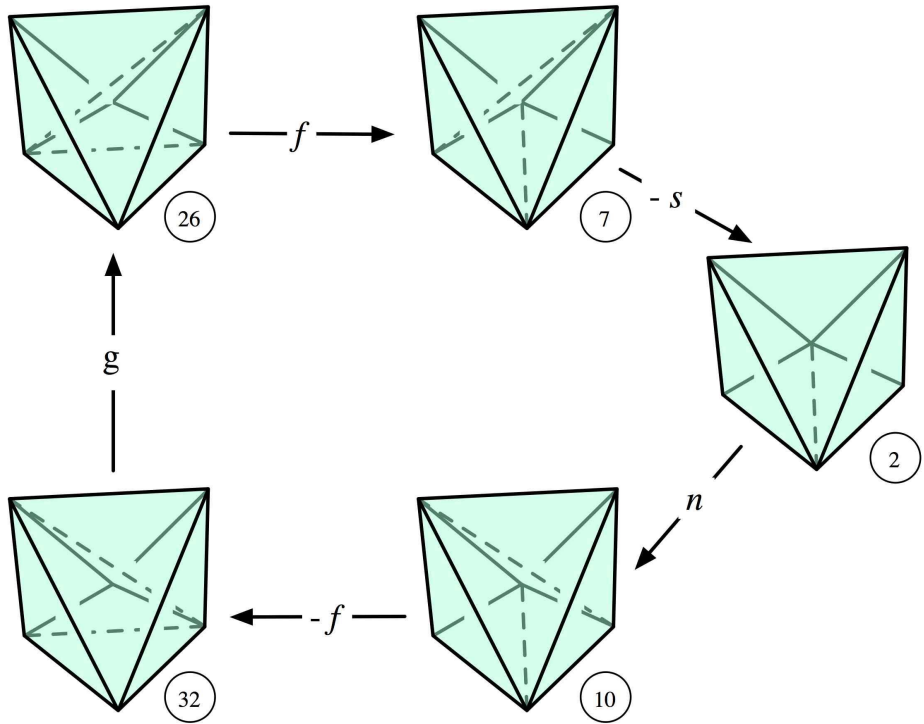}
\caption{The secondary polytope of the genotope in
Figure~\ref{fig:gold}.
It appears as a  face of the secondary polytope of the 3-cube
and its vertices and edges are labeled as in
Table~\ref{tab:74}.   The 
underlying genotype space 
is defined in Example~\ref{threesix} and represents
the HIV fitness landscape shapes that slice off the single
mutants \{M184V\} and \{T215Y\}. }
\label{fig:secgold}
\end{figure}

This suggests studying the fitness landscape on the subset
$\mathcal{G} = \{000$, $110$, $011$, $100$, $101$, $111\}$.
The secondary polytope 
of $\mathcal{G}$ is a pentagon whose vertices
correspond to the shapes~2, 7, 10, 26, and 32, i.e., the
five triangulations of the genotope in Example~\ref{threesix} and 
Figure~\ref{fig:gold}. Figure~\ref{fig:secgold} shows this
pentagon. Vertices and edges are labeled as in
Table~\ref{tab:74}.
The pentagon reveals that shapes~26 and 7,
and 32 and 10, differ only by the circuit
$f$. This circuit represents conditional epistasis between
the two RT loci 184 and 215 in the context of the
protease mutation 90M  
(Table~\ref{tab:hiv2}, Figure~\ref{fig:hiv2}f).
The other edges  are 
the circuits $g$, $n$, and $s$. The circuit $g$
relates the marginal epistasis of the two pairs 
(PRO 90, RT 184) and (PRO 90, RT 215).
The circuit $n$ compares the total two-way epistasis
$E$ to the three-way interaction $u_{111}$, and 
$s$ compares the epistasis in (RT 184, RT 215) to
the remaining pairs plus the three-way interaction.

In summary, analysis of the HIV fitness data has revealed a specific
pentagon in the boundary of the
(four-dimensional) secondary polytope of the 3-cube.
The three dominant shapes 7, 2, and 26 in the 
random fitness landscape are
adjacent on the pentagon, and they correspond to
three of the five triangulations of the genotope in Figure~\ref{fig:gold}.
This geometric characterization of the fitness landscape captures
more adequately the interactions underlying the HIV fitness data.
\par

%%%%%%%%%%%%%%%%%%%%%%%%%%%%%%%%%%%%%%%%%%%%%%%%%%%%%%%%%%%%%%%%%%%%%%%%%
\bigskip
\setcounter{chapter}{7}
\setcounter{equation}{0} %-1
\noindent {\bf 7. Synergistic Epistasis in Drosophila}
\smallskip

Turning to a higher-dimensional genetic system, this section illustrates
our  concepts
and methods with the fitness data for Drosophila melanogaster 
reported by Whitlock and Bourguet (2000). Those authors considered 
five genetic loci, denoted {\tt px/sp}, {\tt b}, {\tt ca}, {\tt e/sr} and {\tt h},
for which they created a set of $32=2^5$ homozygous lines fixed for
all the possible different combinations of mutations. 
We characterize the shape of the fitness landscape they measured,
and we discuss their statistical analysis in light of our findings.
Here, $l=2$ and the genotype space is $\,\mathcal{G} = \{0,1\}^5$.
The binary strings $g\in\mathcal{G}$
represent the subsets of
$\bigl\{{\tt px/sp}, {\tt b}, {\tt ca}, {\tt e/sr}, {\tt h}\bigr\}$.
For instance,  $01011$ represents the mutant
${\tt b}/{\tt e/sr}/{\tt h}$.
In what follows, we list the elements of $\mathcal{G}$ in the
following order:
$$ \begin{matrix} 00000 & 10000 &    01000 & 00100 & 00010 & 00001 &
      11000 & 10100 \\ 10010 & 10001 &   01100 & 01010 & 01001 & 00110 &
       00101 & 00011 \\ 11100 & 11010 &   11001 & 10110 & 10101 & 10011 &
        01110 & 01101 \\ 01011 & 00111 &11110 & 11101 & 11011 & 10111 & 01111 &  11111
\end{matrix}
$$
The first column in (Whitlock and Bourguet, 2000, Tab.~1)
 consists of the fitness
values $w_{i jklm}$ which measure the
relative reproductive fitness of each strain:
$$ \begin{matrix}-0.232 &   -0.850 & -0.312 & -0.214 & 
                -0.847  & \phantom{-}0.507 & -0.238 & -0.490 \\
                -1.030 &  \phantom{-}0.232 & -0.968 & -1.338  &
                -0.034  & -1.47 &  -0.739 & 0.2176 \\
         -0.712  & -1.820 & -0.529 & -0.786 &
         -0.195 & -0.641 & -1.945 & -0.047 \\
          \phantom{-}0.0264 & -1.296 &         -2.446 &-1.973 &
          -1.180 & -1.024 & -1.856 &
         -4.560\end{matrix} $$
These values define the fitness landscape 
$w : \mathcal{G} \rightarrow \mathbb{R}$,
e.g., $w_{01100} = -0.968$.

Whitlock and Bourguet's analysis of epistasis
begins with an examination of the ten two-way
interactions. They find that six of the 10 pairs show significant
interaction. In all of them the direction of the effect is negative
(synergistic epistasis). As we show below, each of the two-way
interactions depends on the genotypes at the remaining loci, and a
refined analysis provides more information.

The shape $\,\Pi_\mathcal{G}[w]\,$ of the Drosophila fitness landscape
is a triangulation of the $5$-cube $\Pi_\mathcal{G}$. It consists of 
the $110$ maximal simplices displayed in 
Figure~\ref{fig:simplices}. Although the figure does not convey an
intuitive image of the shape, it precisely characterizes  the interactions.
For verification, we computed the triangulation twice, 
once using the computer algebra software {\tt Macaulay~2} 
(Grayson and Stillman, 1999)
and once using the geometry software 
{\tt Polymake} (Gawrilow and Joswig, 2001),
both in less than one second running time on a standard PC.

\begin{figure}
\footnotesize
$$
\begin{matrix}
 \!\!  \{00000, 00001, 00011, 00100, 01000, 11000\} \{00000, 00001, 00011
, 00100, 10001, 11000\} \\
 \!\!  \{00000, 00010, 00011, 00100, 01000, 11000\} \{00000, 00010, 00011
, 00100, 10110, 11000\} \\
 \!\!  \{00000, 00010, 00011, 10010, 10110, 11000\} \{00000, 00011, 00100
, 10001, 10100, 11000\} \\
 \!\!  \{00000, 00011, 00100, 10100, 10110, 11000\} \{00000, 00011, 10001
, 10010, 10110, 11000\} \\
 \!\!  \{00000, 00011, 10001, 10100, 10110, 11000\} \{00000, 10000, 10001
, 10010, 10100, 11000\} \\
 \!\!  \{00000, 10001, 10010, 10100, 10110, 11000\} \{00001, 00011, 00100
, 00101, 01101, 10101\} \\
 \!\!  \{00001, 00011, 00100, 01000, 01011, 11000\} \{00001, 00011, 00100
, 01011, 01101, 11000\} \\
 \!\!  \{00001, 00011, 00100, 01101, 10001, 10101\} \{00001, 00011, 00100
, 01101, 10001, 11000\} \\
 \!\!  \{00001, 00011, 01011, 01101, 10001, 11000\} \{00001, 00100, 01000
, 01011, 01101, 11000\} \\
 \!\!  \{00001, 01000, 01001, 01011, 01101, 11000\} \{00001, 01001, 01011
, 01101, 10001, 11000\} \\
 \!\!  \{00010, 00011, 00100, 00110, 01011, 10110\} \{00010, 00011, 00100
, 01000, 01011, 10110\} \\
 \!\!  \{00010, 00011, 00100, 01000, 10110, 11000\} \{00010, 00011, 01000
, 01011, 10110, 11000\} \\
 \!\!  \{00010, 00011, 01011, 10010, 10110, 11000\} \{00010, 00100, 00110
, 01011, 01110, 10110\} \\
 \!\!  \{00010, 00100, 01000, 01010, 01011, 10110\} \{00010, 00100, 01010
, 01011, 01110, 10110\} \\
 \!\!  \{00010, 01000, 01010, 01011, 10110, 11000\} \{00010, 01010, 01011
, 10010, 10110, 11000\} \\
 \!\!  \{00011, 00100, 00101, 00111, 01101, 10101\} \{00011, 00100, 00110
, 00111, 01101, 10110\} \\
 \!\!  \{00011, 00100, 00110, 01011, 01101, 10110\} \{00011, 00100, 00111
, 01101, 10101, 10111\} \\
 \!\!  \{00011, 00100, 00111, 01101, 10110, 10111\} \{00011, 00100, 01000
, 01011, 10110, 11000\} \\
 \!\!  \{00011, 00100, 01011, 01101, 10110, 11000\} \{00011, 00100, 01101
, 10001, 10100, 10101\} \\
 \!\!  \{00011, 00100, 01101, 10001, 10100, 11000\} \{00011, 00100, 01101
, 10100, 10101, 10110\} \\
 \!\!  \{00011, 00100, 01101, 10100, 10110, 11000\} \{00011, 00100, 01101
, 10101, 10110, 10111\} \\
 \!\!  \{00011, 00110, 00111, 01011, 01101, 10110\} \{00011, 00111, 01011
, 01101, 10110, 10111\} \\
 \!\!  \{00011, 01011, 01101, 10001, 10101, 10110\} \{00011, 01011, 01101
, 10001, 10110, 11000\} \\
 \!\!  \{00011, 01011, 01101, 10101, 10110, 10111\} \{00011, 01011, 10001
, 10011, 10110, 10111\} \\
 \!\!  \{00011, 01011, 10001, 10011, 10110, 11000\} \{00011, 01011, 10001
, 10101, 10110, 10111\} \\
 \!\!  \{00011, 01011, 10010, 10011, 10110, 11000\} \{00011, 01101, 10001
, 10100, 10101, 10110\} \\
 \!\!  \{00011, 01101, 10001, 10100, 10110, 11000\} \{00011, 10001, 10010
, 10011, 10110, 11000\} \\
 \!\!  \{00100, 00110, 01011, 01101, 01110, 10110\} \{00100, 01000, 01010
, 01011, 01110, 10110\} \\
 \!\!  \{00100, 01000, 01011, 01100, 01101, 11100\} \{00100, 01000, 01011
, 01100, 01110, 10110\} \\
 \!\!  \{00100, 01000, 01011, 01100, 10110, 11100\} \{00100, 01000, 01011
, 01101, 11000, 11100\} \\
 \!\!  \{00100, 01000, 01011, 10110, 11000, 11100\} \{00100, 01011, 01100
, 01101, 01110, 10110\} \\
 \!\!  \{00100, 01011, 01100, 01101, 10110, 11100\} \{00100, 01011, 01101
, 10110, 11000, 11100\} \\
 \!\!  \{00100, 01101, 10100, 10110, 11000, 11100\} \{00110, 00111, 01011
, 01101, 01110, 10110\} \\
 \!\!  \{00111, 01011, 01101, 01110, 01111, 10110\} \{00111, 01011, 01101
, 01111, 10110, 10111\} \\
 \!\!  \{01000, 01010, 01011, 01110, 10110, 11100\} \{01000, 01010, 01011
, 10110, 11000, 11100\} \\
 \!\!  \{01000, 01011, 01100, 01110, 10110, 11100\} \{01001, 01011, 01101
, 10001, 11000, 11001\} \\
 \!\!  \{01010, 01011, 01110, 10110, 11010, 11100\} \{01010, 01011, 10010
, 10110, 11000, 11010\} \\
 \!\!  \{01010, 01011, 10110, 11000, 11010, 11100\} \{01011, 01100, 01101
, 01110, 10110, 11100\} \\
 \!\!  \{01011, 01101, 01110, 01111, 10110, 11110\} \{01011, 01101, 01110
, 10110, 11100, 11110\} \\
 \!\!  \{01011, 01101, 01111, 10110, 10111, 11110\} \{01011, 01101, 01111
, 10111, 11011, 11110\} \\
 \!\!  \{01011, 01101, 10001, 10101, 10110, 11000\} \{01011, 01101, 10001
, 10101, 11000, 11001\} \\
 \!\!  \{01011, 01101, 10101, 10110, 10111, 11100\} \{01011, 01101, 10101
, 10110, 11000, 11100\} \\
 \!\!  \{01011, 01101, 10101, 10111, 11011, 11100\} \{01011, 01101, 10101
, 11000, 11001, 11100\} \\
 \!\!  \{01011, 01101, 10101, 11001, 11011, 11100\} \{01011, 01101, 10110
, 10111, 11100, 11110\} \\
 \!\!  \{01011, 01101, 10111, 11011, 11100, 11110\} \{01011, 01110, 10110
, 11010, 11100, 11110\} \\
 \!\!  \{01011, 10001, 10011, 10110, 10111, 11011\} \{01011, 10001, 10011
, 10110, 11000, 11011\} \\
 \!\!  \{01011, 10001, 10101, 10110, 10111, 11011\} \{01011, 10001, 10101
, 10110, 11000, 11011\} \\
 \!\!  \{01011, 10001, 10101, 11000, 11001, 11011\} \{01011, 10010, 10011
, 10110, 11000, 11011\} \\
 \!\!  \{01011, 10010, 10110, 11000, 11010, 11011\} \{01011, 10101, 10110
, 10111, 11011, 11100\} \\
 \!\!  \{01011, 10101, 10110, 11000, 11011, 11100\} \{01011, 10101, 11000
, 11001, 11011, 11100\} \\
 \!\!  \{01011, 10110, 10111, 11011, 11100, 11110\} \{01011, 10110, 11000
, 11010, 11011, 11100\} \\
 \!\!  \{01011, 10110, 11010, 11011, 11100, 11110\} \{01101, 01111, 10111
, 11011, 11101, 11110\} \\
 \!\!  \{01101, 10001, 10100, 10101, 10110, 11000\} \{01101, 10100, 10101
, 10110, 11000, 11100\} \\
 \!\!  \{01101, 10101, 10111, 11011, 11100, 11101\} \{01101, 10101, 11001
, 11011, 11100, 11101\} \\
 \!\!  \{01101, 10111, 11011, 11100, 11101, 11110\} \{01111, 10111, 11011, 11101, 11110, 11111\} 
\end{matrix}
$$
\caption{The shape of the Drosophila fitness landscape.}
\label{fig:simplices}
\end{figure}

Returning to the question of synergistic epistasis, we examine
each pair of loci when we fix the values ($0$ or $1$) at 
the remaining three loci.
For example,  the first pair of loci $(1,2)$ (i.e., ({\tt px/sp}, {\tt b}))
has positive epistasis twice, namely when there is either no other mutation 
or only the mutation $3$ ({\tt ac}). However, as soon as either of
mutations $4$ ({\tt e/sr}) or $5$ ({\tt h})
occurs, the epistasis between 
{\tt px/sp} and {\tt b}
becomes negative.  Algebraically, if we write
$$
\alpha_{**klm} \quad = \quad
w_{00klm} + w_{11klm} - w_{01klm} - w_{10klm}
$$
then
$\alpha_{**000} = 0.692 $ and
$\alpha_{**100} = 0.532$ are positive while the
other six epistatic interactions have negative numerical values:
$$ \begin{matrix} 
\alpha_{**001} = -0.220  &&
\alpha_{**010} = -0.299 &&
\alpha_{**011} =  -0.3478 \\
\alpha_{**101} = -2.470 & &
\alpha_{**110} = -1.185 &&
\alpha_{**111} = -2.976.
\end{matrix}
$$
Biologically, this analysis confirms the negative total marginal two-way
epistasis described in (Whitlock and Bourguet, 2000), however it also reveals 
precise information about the conditional epistasis.
Geometrically, an analysis of all two-way interactions 
involves examining the $80$ two-dimensional faces
of the $5$-cube. Of these $80$ squares, we find that
precisely $26$ have positive epistasis. 
We list the numerical values of the $26$ positive interactions
grouped by pairs of loci:
\begin{itemize}
\item[(1,2)]
$\alpha_{**000} = 0.692,\,
\alpha_{* * 1 0 0} = 0.532$,
\item[(1,3)] $
\alpha_{* 0 * 0 0} = 0.342,\,
\alpha_{* 0 * 0 1} = 0.819,\,
\alpha_{* 0 * 1 0} = 0.867,\,
\alpha_{* 0 * 1 1} = 1.1306,\,$ $
\alpha_{* 1 * 0 0} = 0.182,$
\item[(1,4)] $
\alpha_{* 0 0 * 0} = 0.435,\,
\alpha_{* 0 1 * 0} = 0.960, $
\item[(1,5)] $
\alpha_{* 0 0 0 *} = 0.343,\,
\alpha_{* 0 1 0 *} = 0.820, $
\item[(2,3)] $
\alpha_{0 * * 0 1} = 1.233,\,
\alpha_{0 * * 1 0} = 0.016, $
\item[(2,4)] $
\alpha_{0 * 0 * 1} = 0.3498,\,
\alpha_{0 * 1 * 0} = 0.279,\,
\alpha_{1 * 0 * 1} = 0.222, $
\item[(2,5)] $
\alpha_{0 * 0 1 *} = 0.2998,\,
\alpha_{0 * 1 0 *} = 1.446,\,
\alpha_{1 * 0 1 *} = 0.251, $
\item[(3,4)] $
\alpha_{0 1 * * 0} = 0.049,\,
\alpha_{1 0 * * 1} = 0.044, $
\item[(3,5)] $
\alpha_{0 1 * 0 *} = 0.643,$
\item[(4,5)] $
\alpha_{0 0 0 * *} = 0.3256,\,
\alpha_{0 0 1 * *} = 0.699,\,
\alpha_{0 1 0 * *} = 1.0864,\,
\alpha_{1 1 0 * *} = 0.931.$
\end{itemize}
Considering that $54/80=0.675$ is higher than the fraction 0.6 of
negative two-way interactions identified by Whitlock and Bourguet, 
and that many of the positive epistatic interactions are very
small, one can argue a stronger case for
synergistic epistasis.
Importantly, our analysis also reveals exactly 
which pairs of loci can have positive or negative epistasis, and how
this depends on the patterns at the other loci.
The pair $(1,3)$ (i.e., ({\tt px/sp}, {\tt ca}))
has the largest number of positive epistasis patterns, namely five,
while the pair $(3,5)$ (i.e., ({\tt ca}, {\tt h})) has only one. In
order to assess the statistical significance of these interactions,
it is imperative to have replicates of the fitness measurements
(cf.\ Section~6).

The shape of the fitness landscape also reveals
information about the fittest population. To see this, we return to the
geometry in Figure \ref{fig:simplices}.
The polyhedral complex dual to the triangulation,
is referred to as the {\em tight span} (Grier et al., 2006). 
For our data, the tight span is
three-dimensional, with f-vector
$(110,214,127,22)$.  Besides the $22$ three-dimensional
cells, the tight span of $\Pi_{\mathcal{G}}[w]$ has ten maximal cells of dimension
$2$ and two maximal cells of dimension one.
These two ``tentacles'' correspond to  the
genotypes $10000$ and $11111$ which 
are ``sliced off'' in the triangulation, i.e., each of them
lies in only one maximal simplex.
The GKZ vector of the triangulation $\Pi_\mathcal{G}[w]$ equals
$1/120$ times
$$
\begin{matrix}
(12, &1, & 21, & 46, && 13,& 13,& 64,& 17, \\
\, \, 12, &35, &7, &10,&& 3, &7, &2, & 51, \\
\,\, 33, &7,& 7, & 83,&& 30,& 7,& 15,& 64, \\
 \,\, 80,& 9,& 12,& 5,&& 24,& 22,& 7,& \, 1\, )
\end{matrix}
$$
Recall that the entry of this vector indexed by genotope  $g\in \mathcal{G}$
is the probability  that $g$ appears in the
fittest population for a randomly chosen allele frequency vector.
 For instance, for the
strain $10110 = {\tt px}/{\tt sp}/{\tt ca}/{\tt e}/{\tt sr}$
that probability is $\, 83/120 = 69.2 \%$.
The triangulation  $\Pi_{\mathcal{G}}[w]$ can also be written in a way
which is complementary to the list of $110$ maximal simplices.
Namely, there are 332 minimal non-faces of $\Pi_{\mathcal{G}}[w]$,
31 non-triangles, and the one non-tetrahedron
\[
   \{00100,\, 00010,\, 11000,\, 01011\}.
\]
The shape of the fitness landscape implies that these four Drosophila
mutants cannot coexist in a maximally fit population, however
any three of them can.
%These four Drosophila mutants cannot coexist in a fittest population,
%given the shape of the fitness landscape, but any three of them can.
\par

%%%%%%%%%%%%%%%%%%%%%%%%%%%%%%%%%%%%%%%%%%%%%%%%%%%%%%%%%%%%%%%%%%%%%%%%%
\bigskip
\setcounter{chapter}{8}
\setcounter{equation}{0} %-1
\noindent {\bf 8. Discussion}
\smallskip

Our description of the shape of fitness landscapes in terms of
triangulations of the genotope directly reveals information about all the
interactions among the genotypes. In the case of many
organisms, including humans, polymorphism occurs at single nucleotides 
in the genome (SNPs), and is usually of two
types. Thus, even though humans are diploid, and in principle there
could be 16 possible alleles at a polymorphic site, there are usually
only $l = 2$. It is important to note
that even though the number of SNPs is in the millions, 
the {\em human genotope} is determined
only by the genotypes occurring in the population. The linkage 
disequilibrium structure of the human population
(The International HapMap Consortium, 2005) 
suggests that the dimension of 
this genotope is far smaller than would be suggested by the number of
SNPs. Thus, there is hope that in the future,
with measurements of fitness one can learn about 
populations by examining the shapes of fitness landscapes
on the human genotope. In the
meantime, the mathematics we have developed will be useful for studying
interactions among genotypes in small mutation studies.

Another interpretation of the triangulations of the genotope 
is in terms of the genotypes that can occur
in maximally fit populations. Such populations must consist of genotypes 
that label one simplex in the triangulation. In other words, the
description of the shape of fitness landscapes that we have provided 
is fundamental for understanding the genotypes of populations that
evolve by recombination but without mutation.
In the case of populations that evolve
with mutation, but without recombination, a complementary analysis of
fitness landscapes in terms of linear extensions of the genotope (viewed
as a poset) is provided by Weinreich (2005). 
An understanding of the
relationship between these approaches will lead to a deeper
understanding of populations that evolve by recombination and mutation.

What our study and Weinreich's have
in common is that they fall into the domain of non-parametric statistics.
In contrast to other papers on fitness landscapes,
including (Karlin and Feldman, 1970) and 
(Kondrashov and Kondrashov, 2001),
we do not make {\em a priori} assumptions
about the fitness landscape. In particular, our analysis of the data
in Sections 6 and 7 is not based on any choice of model for $w$.
On the other hand, the geometry of the secondary polytope interfaces
well with  Bayesian statistics, because any family of distributions on
$\mathbb{R}^\mathcal{G}$ induces a family of distributions
on the finite set of possible shapes.
\par

%%%%%%%%%%%%%%%%%%%%%%%%%%%%%%%%%%%%%%%%%%%%%%%%%%%%%%%%%%%%%%%%%%%%%%%%%
\bigskip
\noindent {\large\bf Acknowledgments}
\smallskip

\noindent
This paper was inspired by the DARPA workshop 
on Fitness Landscapes held at Berkeley in February 2006. 
We thank 
Sebastian Bonhoeffer, Richard Lenski and Sally Otto for helpful discussions.
N.B.\ was supported by
the Deutsche Forschungsgemeinschaft (BE~3217/1-1). L.P.\ and B.S.\ 
were  supported by the DARPA program
``Fundamental Laws in Biology'' (HR0011-05-1-0057).
\par

%%%%%%%%%%%%%%%%%%%%%%%%%%%%%%%%%%%%%%%%%%%%%%%%%%%%%%%%%%%%%%%%%%%%%%%%%
\bigskip
\noindent{\large\bf References}
%\smallskip

\begin{description}
\setlength{\itemsep}{0cm} \setlength{\parsep}{0cm}

\item 
Bateson, W. (1909).
{\em Mendel's Principles of Heredity}.
Cambridge University Press, Cambridge, UK.

\item
Beerenwinkel, N., Eriksson, N. and Sturmfels, B. (2006).
Evolution on distributive lattices.
{\em J. Theor. Biol., to appear.}
%\newblock URL \url{http://arxiv.org/abs/q-bio/0511039}.

\item
Bonhoeffer, S., Chappey, C., Parkin, N.~T., Whitcomb, J.~M. 
  and Petropoulos, C.~J. (2004).
Evidence for positive epistasis in {HIV}-1.
{\em Science} {\bf 306(5701)}, 1547--1550.
%\newblock URL \url{http://dx.doi.org/10.1126/science.1101786}.

\item
Cordell, H.~J. (2002).
Epistasis: what it means, what it doesn't mean, and statistical
  methods to detect it in humans.
{\em Hum. Mol. Genet.} {\bf 11(20)}, 2463--2468.

\item
De Loera, J., Rambau, J. and Santos, F. (2006).
{\em Triangulations: Applications, Structures, and Algorithms}.
%Algorithms and Computation in Mathematics. 
Springer, 
{\em to appear}.

\item
Feldman, M.~W., Franklin, I. and Thomson, G.~J. (1974).
Selection in complex genetic systems. {I}. {T}he symmetric equilibria
  of the three-locus symmetric viability model.
{\em Genetics} {\bf 76(1)}, 135--162.

\item
Fisher, R.~A. (1918).
The correlations between relatives on the supposition of {M}endelian
  inheritance.
{\em Trans. R. Soc. Edinburgh} {\bf 52}, 399--433.

\item
Gavrilets, S. (2004).
{\em Fitness Landscapes and the Origin of Species.} 
%volume~41 of {\em Monographs in Population Biology}.
Princeton University Press.

\item
Gawrilow, E. and Joswig, M. (2001).
Polymake: an approach to modular software design in computational geometry.
{\em Proc. 17th ACM Symposium on Computational Geometry}, Medford, MA.

\item
Grayson, D.~R. and Stillman, M.~E. (1999).
Macaulay~2, a software system for research in algebraic geometry.
\url{http://www.math.uiuc.edu/Macaulay2/}.

\item
Grier, D., Huggins, P., Sturmfels, B. and Yu, J. (2006).
The hyperdeterminant and triangulations of the 4-cube.
{\em submitted}.

\item
Hallgr\'{\i}msd\'ottir, I.~B. (2005).
Statistical methods for gene mapping in complex diseases.
Ph.D. thesis in Statistics, UC Berkeley.

\item
Hendy. M.~D. and Charleston, M.A. (1993).
Hadamard conjugation: a versatile tool for modelling nucleotide
  sequence evolution.
{\em New Zealand J. Botany} {\bf 31}, 231--237.

\item
Karlin, S. and Feldman, M.~W. (1970).
Linkage and selection: two locus symmetric viability model.
{\em Theor. Popul. Biol.} {\bf 1(1)}, 39--71.

\item
Kondrashov, F.~A. and Kondrashov, A.~S. (2001).
Multidimensional epistasis and the disadvantage of sex.
{\em Proc. Natl. Acad. Sci. U.S.A.} {\bf 98(21)}, 12089--12092.
%\newblock URL \url{http://dx.doi.org/10.1073/pnas.211214298}.

\item
Lindman, H.~R. (1974).
{\em Analysis of Variance in Complex Experimental Designs.}
Freeman and Co., San Francisco, CA.

%\item
%Miller, R.~G. (1997).
%{\em Beyond ANOVA: Basics of Applied Statistics.}
%Chapman \& Hall, Boca Raton, FL. 

\item
Phillips, P.~C. (1998).
The language of gene interaction.
{\em Genetics} {\bf 149(3)}, 1167--1171.

\item
Rambau, J. (2002).
TOPCOM: Triangulations of point configurations and oriented matroids.
{\em Proc. Int. Congress of Mathematical Software ICMS.}

\item
Sanju\'{a}n, R., Moya, A. and Elena, S.~F. (2004).
The contribution of epistasis to the architecture of fitness in an RNA virus.
{\em Proc. Natl. Acad. Sci. U.S.A.} {\bf 101(43)}, 15376--15379.
%\newblock URL \url{http://dx.doi.org/10.1073/pnas.0404125101}.

\item
Segal, M.~R., Barbour, J.~D. and Grant, R.~M. (2004).
Relating HIV-1 sequence variation to replication capacity via trees and forests.
{\em Stat. Appl. Genet. Mol. Biol.} {\bf 3(1)}, 2.

\item
The International HapMap Consortium. (2005).
A haplotype map of the human genome.
{\em Nature} {\bf 437(7063)}, 1299--1320.

\item
Wainberg, M.~A. (2004).
The impact of the M184V substitution on drug resistance and viral fitness.
{\em Expert Rev. Anti. Infect. Ther.} {\bf 2(1)}, 147--151.

\item
Weinreich, D.~M. (2005).
The rank ordering of genotypic fitness values predicts genetic
  constraint on natural selection on landscapes lacking sign epistasis.
{\em Genetics} {\bf 171(3)}, 1397--1405.
%\newblock URL \url{http://dx.doi.org/10.1534/genetics.104.036830}.

\item
Whitlock, M.~C. and Bourguet, D. (2001).
Factors affecting the genetic load in {D}rosophila: synergistic
  epistasis and correlations among fitness components.
{\em Evolution Int. J. Org. Evolution} {\bf 54(5)}, 1654--1660.

\item
Wright, S. (1931).
Evolution in Mendelian populations.
{\em Genetics} {\bf 16}, 97--159.

\item
Ziegler, G.~(1995).
{\em Lectures on Polytopes}.
%Graduate Texts in Mathematics, 152.
Springer, New York, NY.
\end{description}

\vskip .65cm
\noindent
Department of Mathematics, University of California, Berkeley, CA 94720
\vskip 2pt
\noindent
E-mail: \url{{niko,lpachter,bernd}@math.berkeley.edu}
%\vskip 2pt
%\noindent
%second author affiliation
%\vskip 2pt
%\noindent
%E-mail: (second author email)
\vskip .3cm
%\centerline{(Received xxx 200?; accepted xxx 200?)}\par
\end{document}